\documentclass[longauth]{aa}  

\usepackage{graphicx}
\usepackage{txfonts}
\usepackage{amssymb}
\usepackage{multirow}
\usepackage{float}
\usepackage{gensymb}
\usepackage{xcolor}
\usepackage{rotating}
\usepackage[]{hyperref}
\hypersetup{backref=true, pagebackref=true, hyperindex=true, breaklinks=true,colorlinks=true,urlcolor=blue, linkcolor=blue, citecolor=blue,pagecolor=red, bookmarks=true, bookmarksopen=true}

\begin{document}

   \title{The SPHERE view of the jet and the envelope of RY Tau \thanks{Based on observations performed with VLT/SPHERE under program ID 096.C-0241(C), 096.C-0241(F), 096.C-0248(A), 096.C-0454(A), and 1100.C-0481(A).}}

   \subtitle{}

   \author{A.\,Garufi \inst{\ref{Firenze}}
    \and L.\,Podio \inst{\ref{Firenze}}
    \and F.\,Bacciotti \inst{\ref{Firenze}}
    \and S.\,Antoniucci \inst{\ref{Roma}}
    \and A.\,Boccaletti\inst{\ref{LESIA}}
    \and C.\,Codella\inst{\ref{Firenze}, \ref{IPAG}}
    \and C.\,Dougados\inst{\ref{IPAG}}
    \and F.\,M\'{e}nard\inst{\ref{IPAG}}
    \and \\ D.\,Mesa\inst{\ref{Padova}}
    \and M.\,Meyer\inst{\ref{Michigan}}
    \and B.\,Nisini\inst{\ref{Roma}}
    \and H.M.\,Schmid\inst{\ref{ETH}}
    \and T.\,Stolker\inst{\ref{ETH}}
    \and J.L.\,Baudino\inst{\ref{Oxford}}
    \and B.\,Biller\inst{\ref{Edinburgh}}
    \and M.\,Bonavita\inst{\ref{Edinburgh}}
    \and \\ M.\,Bonnefoy\inst{\ref{IPAG}}
    \and F.\,Cantalloube\inst{\ref{MPIA}}
    \and G.\,Chauvin\inst{\ref{IPAG}}
    \and A.\,Cheetham\inst{\ref{Geneva}}
    \and S.\,Desidera\inst{\ref{Padova}}
    \and V.\,D'Orazi\inst{\ref{Padova}}
    \and M.\,Feldt\inst{\ref{MPIA}}
    \and \\ R.\,Galicher\inst{\ref{LESIA}}
    \and A.\,Grandjean\inst{\ref{IPAG}}
    \and R.\,Gratton\inst{\ref{Padova}}
    \and J.\,Hagelberg\inst{\ref{ETH}}
    \and A.M.\,Lagrange\inst{\ref{IPAG}}
    \and M.\,Langlois\inst{\ref{CRAL}}
    \and J.\,Lannier\inst{\ref{IPAG}}
    \and \\ C.\,Lazzoni\inst{\ref{Padova2}}
    \and A.L.\,Maire\inst{\ref{Liege}}
    \and C.\,Perrot\inst{\ref{LESIA}, \ref{Valpo1}, \ref{Valpo2}}
    \and E.\,Rickman\inst{\ref{Geneva}}
    \and T.\,Schmidt\inst{\ref{LESIA}}
    \and A.\,Vigan\inst{\ref{Marseille}}
    \and A.\,Zurlo\inst{\ref{Marseille}, \ref{SANTIAGO}, \ref{SANTIAGO2}}
    \and A. Delboulb\'e\inst{\ref{IPAG}}
    \and \\ D. Le Mignant\inst{\ref{Marseille}} 
    \and D. Fantinel\inst{\ref{Padova}}
    \and O. M\"oller-Nilsson\inst{\ref{MPIA}} 
    \and L. Weber\inst{\ref{Geneva}}
    \and J.-F. Sauvage\inst{\ref{Marseille}}
     }

\institute{INAF, Osservatorio Astrofisico di Arcetri, Largo Enrico Fermi 5, I-50125 Firenze, Italy. \label{Firenze}
  \email{agarufi@arcetri.astro.it}  
  \and INAF - Osservatorio Astronomico di Roma, Via Frascati 33, 00078, Monte Porzio Catone (RM), Italy \label{Roma}
   \and LESIA, Observatoire de Paris, Universit\'e PSL, CNRS, Sorbonne Universit\'e, Univ. Paris Diderot, Sorbonne Paris Cit\'e, 5 place Jules Janssen, 92195 Meudon, France \label{LESIA} 
    \and Univ. Grenoble Alpes, CNRS, IPAG, F-38000 Grenoble, France \label{IPAG}
      \and INAF - Osservatorio Astronomico di Padova, Vicolo dell'Osservatorio 5, 35122 Padova, Italy \label{Padova}    
      \and Department of Astronomy, University of Michigan, 1085 S.\,University Ave, Ann Arbor, MI 48109-1107, USA \label{Michigan}
    \and Institute for Particle Physics and Astrophysics, ETH Zurich, Wolfgang-Pauli-Strasse 27, 8093 Zurich, Switzerland \label{ETH} 
    \and Department of Physics, University of Oxford, Oxford, UK \label{Oxford}
    \and Institute for Astronomy, University of Edinburgh, EH9 3HJ, Edinburgh, UK \label{Edinburgh}
    \and Max Planck Institute for Astronomy, K\"{o}nigstuhl 17, 69117 Heidelberg, Germany \label{MPIA} 
      \and Geneva Observatory, University of Geneva, Ch. des Maillettes 51, 1290, Versoix, Switzerland \label{Geneva} 
      \and CRAL, CNRS, Universit\'{e} Lyon 1, 9 avenue Charles Andr\'{e}, 69561 Saint Genis Laval Cedex, France \label{CRAL}
       \and Universit\`{a} degli Studi di Padova, dipartimento di Fisica e Astronomia, vicolo dell'osservatorio 3, 35122 Padova \label{Padova2}           
       \and STAR Institute, University of Li\`{e}ge, All\'{e}e du Six Ao\^{u}t 19c, B-4000 Li\`{e}ge, Belgium \label{Liege}
       \and Instituto de F\'isica y Astronom\'ia, Facultad de Ciencias, Universidad de Valpara\'iso, Av. Gran Breta\~na 1111, Valpara\'iso, Chile \label{Valpo1}
       \and N\'ucleo Milenio Formaci\'on Planetaria - NPF, Universidad de Valpara\'iso, Av. Gran Breta\~na 1111, Valpara\'iso, Chile \label{Valpo2}
      \and Aix Marseille Universit\'{e}, CNRS, LAM - Laboratoire d'Astrophysique de Marseille, UMR 7326, 13388, Marseille, France \label{Marseille}
  \and N\'ucleo de Astronom\'ia, Facultad de Ingenier\'ia y Ciencias, Universidad Diego Portales, Av. Ejercito 441, Santiago, Chile \label{SANTIAGO} 
  \and Escuela de Ingenier\'ia Industrial, Facultad de Ingenier\'ia y Ciencias, Universidad Diego Portales, Av. Ejercito 441, Santiago, Chile \label{SANTIAGO2}
             }

   \date{Received -; accepted -}

 
 \abstract{Jets are rarely associated with pre-main-sequence intermediate-mass stars. This contrasts with the frequent detection of jets in lower-mass or younger stars. Optical and near-IR observations of jet-driving sources are often hindered by the presence of a natal envelope.}{Jets around partly embedded sources are a useful diagnostic to constrain the geometry of the concealed protoplanetary disk. In fact, the jet-driving mechanisms are affected by both spatial anisotropies and episodic variations at the (sub-)au scale from the star.}{We obtained a rich set of high-contrast VLT/SPHERE observations from 0.6 $\mu$m to 2.2 $\mu$m of the young intermediate-mass star RY Tau. Given the proximity to the Sun of this source, our images have the highest spatial resolution ever obtained for an atomic jet (down to $\sim$4 au).}{Optical observations in polarized light show no sign of the protoplanetary disk detected by ALMA. Instead, we observed a diffuse signal resembling a remnant envelope with an outflow cavity. The jet is detected in the H$\alpha$, [S \textsc{ii}] at 1.03 $\mu$m, He \textsc{i} at 1.08 $\mu$m, and [Fe \textsc{ii}] lines in the 1.25 $\mu$m and 1.64 $\mu$m. The jet appears to be wiggling and its radial width increasing with the distance is complementary to the shape of the outflow cavity suggesting a strong jet/envelope interaction. Through the estimated tangential velocity ($\approx$ 100 km s$^{-1}$), we revealed a {possible} connection between the launching time of the jet {sub-structures} and the stellar activity of RY Tau.}{RY Tau is at an intermediate stage toward the dispersal of the natal envelope. This source shows episodic increases of mass accretion/ejection similarly to other known intermediate-mass stars. The amount of observed jet wiggle is consistent with the presence of a precessing disk warp or misaligned inner disk that would be induced by an unseen planetary/sub-stellar companion at sub-/few-au scales. The high disk mass of RY Tau and of two other jet-driving intermediate-mass stars, HD163296 and MWC480, suggests that massive, full disks are more efficient at launching prominent jets.}

   \keywords{stars: pre-main sequence --
                protoplanetary disks --
                ISM: individual object: RY Tau --
                ISM: jets and outflows
                }

\authorrunning{Garufi et al.}

\titlerunning{The SPHERE view of RY Tau}

   \maketitle
%

\section{Introduction}
The formation history of planets is tightly connected to the global evolution of protoplanetary disks. Our knowledge of the disk morphology has greatly advanced in the last few years thanks to the high-contrast and high-resolution imaging in both the visible/near-IR (NIR) and the (sub-)mm offered by new instruments like VLT/SPHERE, GPI, and ALMA. However, if on one hand ALMA has also access to younger, embedded sources \citep[see e.g.,][]{Sheehan2017}, the optical/NIR imaging has important limitations in this regard because of their much higher extinction. In fact, the protoplanetary disk of well studied pre-main sequence (PMS) stars like e.g., HL Tau {\citep{Takami2007}} or DG Tau B \citep{Podio2011} have not been imaged at short wavelengths, and the whole sample of disks with available NIR images is biased toward more evolved sources with no remnant envelope \citep{Garufi2018}.   

Among the various physical processes that affects the morphology of protoplanetary disks, the removal of mass and angular momentum through jets and outflows {may play} a fundamental role \citep[see the review by][]{Frank2014} since it {can limit} the increase of stellar rotation and allows high accretion rate for a large fraction of the protoplanetary disk lifetime. Given the importance of this element for the star formation process, it is crucial to establish if the ejection of jets is a general phenomenon for forming stars of all masses and evolutionary stages. Jets are also a useful diagnostic of the innermost disk regions being tightly connected to the accretion processes and the star/disk interplay.

\begin{table*}
\centering
\caption{Summary of observations. Columns are: instrument, observing mode, date, waveband, exposure time, angular resolution, main target emission, and relative figure in this work. DBI: dual band imaging, IFS: integral field spectroscopy, DPI: differential polarization imaging, RDI: reference-star differential imaging, SDI: spectral differential imaging.}
\label{Observing_log}
\begin{tabular}{cccccccc}
\hline
Instrument & Mode & Waveband & Date & $t_{\rm exp}$ (min) & $\theta$ (mas) & Main target emission & Figure \\
\hline
\hline
\multirow{4}{*}{IRDIS} & \multirow{4}{*}{DBI} & $H23$ & 2015-10-27 & 68 & 60 & Continuum, [Fe \textsc{ii}] line & \ref{Jet_Imagery} (top panels), \ref{IRDIS_Jet}, \ref{Collage} \\
& & $K12$ & 2015-12-26 & 51 & 80 & Continuum, H$_2$ line & - \\
& & $K12$ & 2016-01-01  & 85 & 80 & Continuum, H$_2$ line & - \\
& & $H23$ & 2017-11-29 & 102 & 60 & Continuum, [Fe \textsc{ii}] line & \ref{Jet_Imagery} (top panels), \ref{IRDIS_Jet} \\
\hline
\multirow{4}{*}{IFS} & \multirow{4}{*}{IFS} & $YJ$ & 2015-10-27  & 68 & 40 & [S \textsc{ii}], He \textsc{i}, [Fe \textsc{ii}] lines & - \\
&  & $YH$ & 2015-12-26 & 51 & 50 & [S \textsc{ii}], He \textsc{i}, [Fe \textsc{ii}] lines & - \\
&  & $YH$ & 2016-01-01 & 85 & 50 & [S \textsc{ii}], He \textsc{i}, [Fe \textsc{ii}] lines & - \\
&  & $YJ$ & 2017-11-29 & 102 & 40 & [S \textsc{ii}], He \textsc{i}, [Fe \textsc{ii}] lines & \ref{Jet_Imagery} (bottom left panels) \\
\hline
\multirow{3}{*}{ZIMPOL} & DPI & $I'$ & 2015-12-17  & 40+16 & 30 & Polarized continuum & \ref{Imagery_ZIMPOL}, \ref{Collage} \\
&  RDI & $I'$ & 2015-12-27  & 11+11 & 30 & Continuum & \ref{Imagery_ZIMPOL} \\
& SDI & H$\alpha$ & 2015-11-08  & 47 & 30 & H$\alpha$ & \ref{Jet_Imagery} (bottom right panel), \ref{Collage} \\
\hline
\hline
\end{tabular} 
\end{table*}

It is known that jets are almost ubiquitous in protostars \citep[see e.g.,][]{Nisini2015}. However, jets are often observed in more evolved PMS stars, and in particular around the low-mass ($\lesssim1.5\,\rm M_{\odot}$) TTSs \citep[e.g.,][]{Hartigan1995, Bacciotti2002}. The observations of these stars are facilitated by their intrinsic abundance and by their long PMS timescale that allows the direct imaging of their disk after their parent envelope is dissipated. These low-mass stars preserve their outer convective zones until and beyond the zero-age main sequence (ZAMS), and are thus accompanied by strong magnetic field. On the other hand, intermediate-mass stars ($1.5-5\,\rm M_{\odot}$) develops a radiative envelope throughout their PMS stage \citep{Gregory2012}. Observationally, these stars have initially late spectral type (K to F) and are referred to as intermediate-mass TTSs {\citep[IMTTSs, see e.g.,][]{Calvet2004}} while at a later stage of their PMS evolution they appear as Herbig AeBe stars and eventually turn into A/B main-sequence stars. The magnetic field of Herbig stars is weaker and less characterized \citep{Hubrig2018}. Unlike TTSs, observations of jets around intermediate- and high-mass {young} stars are rare, although a few examples exist \citep[e.g.,][]{Devine2000, Antoniucci2016}.

RY Tau is an excellent laboratory for the study of jets around intermediate-mass stars and of late stages of envelope dispersal. This source is in fact associated to {an extended nebulosity and a remnant envelope close to the star} \citep{Nakajima1995, Takami2013} as well as to an optically bright jet \citep{Cabrit1990, St-Onge2008, Agra-Amboage2009, Coffey2015, Skinner2018}. High-amplitude photospheric variations \citep{Petrov1999} and the primordial spectral energy distribution \citep[SED,][]{Robitaille2007} also points to a relatively young source. Yet, its protoplanetary disk shows rings and an inner cavity in the sub-mm \citep{Isella2010, Long2018b} that are typically observed in more evolved sources.

In this work, we present new optical and NIR observations of RY Tau obtained with VLT/SPHERE \citep{Beuzit2019}. Although this instrument was designed and optimized for the planet and disk characterization, it also allows us to image the inner regions of jets in the atomic lines at an unprecedented resolution of $\approx30$ mas, corresponding to 4$-$5 au at 130 pc. The paper is organized as follows. In Sect.\,\ref{Observations}, we describe the analyzed datasets. Then, in Sect.\,\ref{Stellar_properties} and Sect.\,\ref{RY_Tau} we present our results from spectroscopic, photometric, and imaging data. Finally, in Sect.\,\ref{Discussion} and Sect.\,\ref{Conclusions}, we discuss our findings and conclude.
   
\section{Observations and data reduction} \label{Observations}
\subsection{VLT/SPHERE images} \label{Observations_SPHERE}
Multiple images of RY Tau were taken in the context of the guaranteed time observations of the high-contrast imager SPHERE \citep{Beuzit2019}, operating at the Very Large Telescope (VLT) in conjunction with the extreme adaptive optics (AO) system SAXO \citep{Fusco2006}.

\textit{Differential polarization imaging (DPI).} Polarimetric observations of RY Tau were taken with the sub-instrument ZIMPOL \citep[Zurich imaging polarimeter,][]{Schmid2018} in the $I'$ band ($\lambda_{\rm c}=790$ nm) on 2015 December 17. To minimize instrument artifacts, the observations were performed in DPI field stabilized mode by alternating the object orientation by $60\degree$ and dithering its position by 14 pixels. We operated in both SlowPolarimetry mode, which maximizes the signal-to-noise of the outer regions, and FastPolarimetry mode, which provides the highest polarimetric precision of the inner regions. In SlowPolarimetry mode we adopted a 155 mas-diameter Lyot coronagraph and DIT=50 sec, while in FastPolarimetry a DIT=20 sec and no coronagraph was employed. The DPI data reduction follows the standard technique adopted in several works involving VLT/NACO and SPHERE observations \citep[see e.g.,][]{Avenhaus2014b, Garufi2016} and provides as a final product the Stokes parameters $Q$ and $U$, as well as the polarized intensity $PI=(Q^2+U^2)^{1/2}$ and the radial parameters $Q_{\phi}$ and $U_{\phi}$ \citep{Schmid2006}.

\textit{Reference-star differential imaging (RDI).} ZIMPOL observations of RY Tau in the $I'$ band were also taken in Classical Imaging (CI) on 2015 December 17, along with those of a reference star, HD284071. Observations were taken in pupil stabilized mode with DIT=10 sec. The reduction followed the basic steps of the ZIMPOL/DPI reduction, leading to the extraction of 70 individual intensity frames for both RY Tau and reference star that were scaled, subtracted, and averaged to generate a unique residual image. 

\textit{Spectral differential imaging (SDI).} RY Tau was also observed with ZIMPOL on 2015 November 8 in the H$\alpha$  ($\lambda_{\rm c}=655.6$ nm, $\Delta \lambda=5.5$ nm) and in the adjacent continuum ($\lambda_{\rm c}=644.9$ nm, $\Delta \lambda=4.1$ nm) filters (Program ID 096.C-0454(A), P.I.\ Dougados). Images were acquired in field-stabilized mode with no coronagraph for a total integration time on source of 47 minutes.
The data were processed using the ETH Zurich ZIMPOL IDL pipeline (version 1.5), which applies standard bias, dark, and flat-field corrections, and allows for re-centering, derotation, and combination of the frames. The data reduction followed the same procedure described by \citet{Antoniucci2016}. In particular, we first normalized each continuum frame to the corresponding H$\alpha$ frame to take into account the different filter band-passes, using the ratio of the total counts measured in the star peak region in the two images as normalization factor. Then, we performed a frame-by-frame subtraction of the H$\alpha$ and continuum exposures to remove the continuum emission. All the continuum-subtracted frames thus obtained were eventually de-rotated and averaged to generate the final H$\alpha$ image. 

\textit{Integral field spectroscopy (IFS) and dual band imaging (DBI).} Finally, RY Tau was observed as part of the SpHere INfrared survey for Exoplanets \citep[SHINE,][]{Chauvin2017} four times between 2015 October 27 and 2017 November 29. Observations consist of simultaneous imaging with the infrared dual-band imager and spectrograph \citep[IRDIS,][]{Dohlen2008} and integral field spectroscopy instrument \citep[IFS,][]{Claudi2008}. They were done in both IRDIFS and IRDIFS\_EXT modes, that is with IFS operating between Y and J spectral bands ($\lambda=0.95-1.35\,\mu$m) and IRDIS simultaneously performing DBI \citep{Vigan2010} in H23 band ($\lambda_{\rm c}=1.58/1.67\,\mu$m, {$\Delta \lambda=52/54$ nm}) and with IFS operating between Y and H bands ($\lambda=0.95-1.65\,\mu$m) and IRDIS in DBI in K12 band ($\lambda_{\rm c}=2.10/2.25\,\mu$m, {$\Delta \lambda=102/109$ nm}), respectively. {The spectral resolution is $\sim$50 in the IRDIFS and $\sim$30 in the IRDIFS\_EXT modes}. All observations were pupil stabilized to enable angular differential imaging (ADI) and the star was centered under an apodized pupil Lyot coronagraph \citep{Carbillet2011} of radius $r$\,=\,185\,mas. All IFS and IRDIS data sets were reduced with the SPHERE Data Reduction and Handling (DRH) pipeline \citep{Pavlov2008} in the SPHERE Data Center \citep[DC\footnote{\url{http://sphere.osug.fr/spip.php?rubrique16&lang=en}},][]{Delorme2017}. The fully reduced {IFS} images are combined to four-dimensional data cubes (two spatial, one spectral, and one time dimension) and further processed with the SHINE Specal pipeline \citep{Galicher2018}. The latter performs angular and spectral differential imaging ({ADI+SDI)} in the flavors of classical ADI \citep[cADI,][]{Marois2006}, Template Locally Optimized Combination of Images \citep[TLOCI,][]{Marois2014}, and Principal Component Analysis \citep[PCA,][]{Soummer2012, Amara2015}. {SDI is performed on the IFS images by constructing a median PSF from all available channels and by removing this reference image to each individual channel, thus as to enhance the contrast of line emission compared to continuum emission.}

All SPHERE images have been corrected for the SPHERE True North of $-1.8\degree$ \citep{Maire2016}. The angular resolution achieved is approximately 30 mas for the ZIMPOL images, 40$-$50 mas for the IFS images, and 60$-$80 mas for the IRDIS images. A summary of observations is shown in Table \ref{Observing_log}.

\subsection{WHT/UES spectra} \label{UES}
Constraints on the stellar effective temperature can be obtained from optical high-resolution spectra. In this work, we made use of four spectra of RY Tau that were obtained on 25 Oct 1998 and 28, 29, 31 Jan 1999 with the 4.2-m William Herschel Telescope (WHT) on the Observatorio del Roque de los Muchachos (La Palma, Spain) equipped with the Utrecht Echelle Spectrograph (UES). The observations belong to the International Time program \mbox{EXPORT} \citep{Eiroa2000}. UES was set to provide a wavelength coverage between 3800 and 5900 \AA. The spectra were dispersed into 59 echelle orders with a resolving power of 49\,000. The slit width was set to 1.15\arcsec \ projected on the sky. Details about the reduction of the spectra can be found in \citet{Mora2001}.

 \begin{figure*}
  \centering
 \includegraphics[width=18cm]{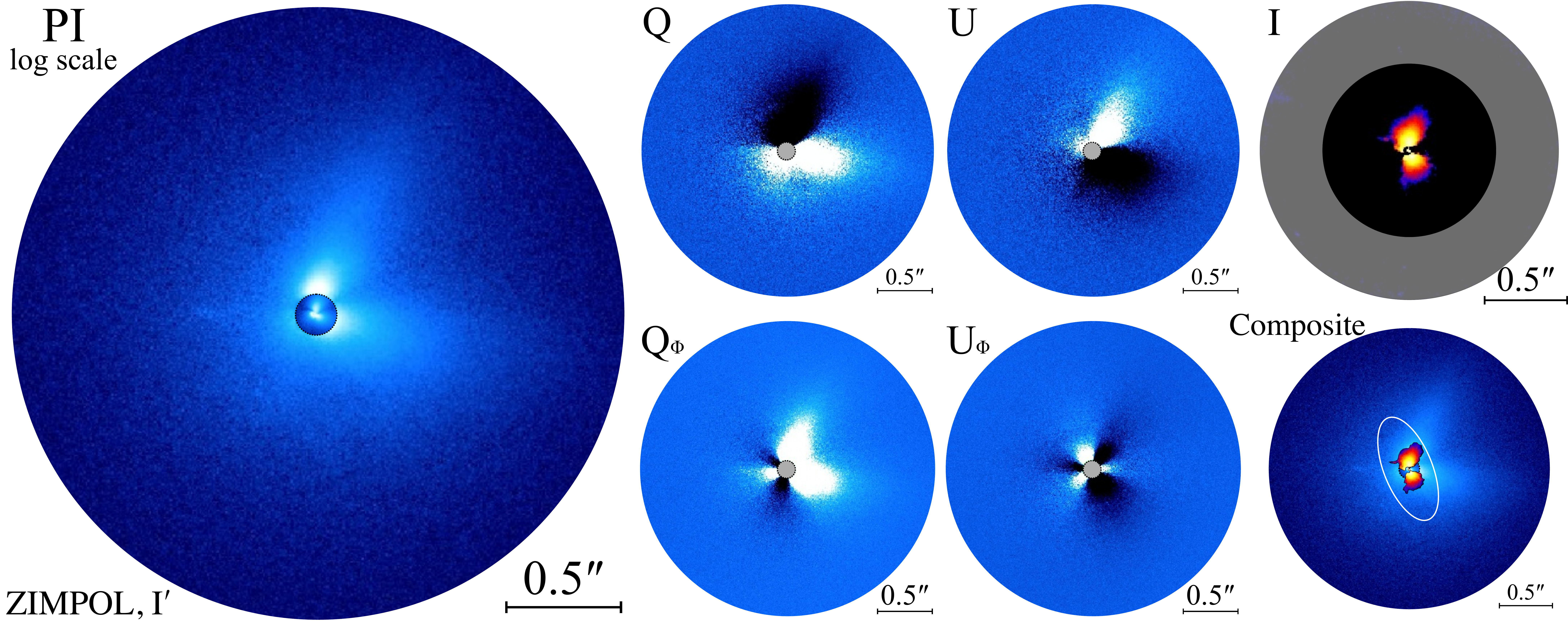} 
      \caption{Imagery of RY Tau in the $I'$ band from ZIMPOL. The polarized intensity $PI$ is shown to the left in logarithmic scale with the main region obtained with coronagraphic {SlowPol} exposures and the small, inner region with {FastPol} exposures (see text). The $Q$, $U$, $Q_{\phi}$, and $U_{\phi}$ parameters are shown in the two central columns with the same linear scale (note the different scale from $PI$) where black regions indicate negative values. The total intensity image obtained in RDI is shown in logarithmic scale to top right {(the black pixels indicate negative values and the grey mask unreliable results)}, along with a graphic comparison with $PI$ to bottom right. The white ellipse outlines the protoplanetary disk seen by ALMA \citep{Long2018b}. North is up, East is left.}
          \label{Imagery_ZIMPOL}
  \end{figure*}

\section{Stellar properties} \label{Stellar_properties}
The stellar properties of RY Tau from the literature are often discrepant. In this section, we perform their new calculation and discuss the critical estimate of the distance to the source. {These parameters are summarized in Table \ref{Table_star}.} 

\subsection{The distance} \label{RYTau_distance}
RY Tau belongs to the Taurus star-forming region \citep{Kenyon1994} and it has been typically assumed to be at $133^{+55}_{-30}$ pc, based on the Hipparcos data \citep{Esa1997}. However, Gaia DR1-TGAS \citep[][]{GAIA2016} revealed $d=176^{+32}_{-23}$ pc and Gaia DR2 \citep[][]{GAIA2018} $d=442^{+53}_{-42}$ pc. The large inconsistency between the Hipparcos and Gaia estimates raised some doubts about the latter. DR2 is in fact still a preliminary release, where several data processing loops are not closed and various systematic effects are still present \citep{Lindegren2018}. 

To test the possibility that the Gaia measurement is erroneous, we look at 29 other members of its (visual) parent cloud, L1495 \citep[see][]{Luhman2018}. All these sources have $d$ between 119.5 and 139.1 pc ($\overline{d}$=129.2 pc) and an average proper motion of 8.97 mas ($\sigma=1.27$) and $-24.97$ ($\sigma=1.26$) mas in right ascension and declination. RY Tau shows a consistent proper motion of 9.08 and $-23.05$ from Hipparcos and of 9.10 and $-25.86$ from Gaia DR1, while the DR2 shows a slightly discrepant value (4.98,$-25.71$).   

Given these premises, we are inclined to consider the Gaia DR2 measurement (442 pc) erroneous and the Gaia DR1 (176 pc) less likely than that from Hipparcos (133 pc) since this estimate also matches the average (129 pc) for the other co-moving members of the parent cloud of RY Tau. An investigation of the miscalculation from Gaia is beyond the scope of this paper. We nonetheless note that this may have a connection with the variability of the Hipparcos astrometric solution \citep{Bertout1999, Agra-Amboage2009} indicating motion of the photocenter or with the presence of a bright envelope, as in R CrA \citep{Mesa2019}.  

\subsection{Stellar temperature, mass, and age} \label{Stellar_properties_RYTau}
Two different estimates of the effective temperature $T_{\rm eff}$ are commonly adopted in the literature, i.e.\ $T_{\rm eff}=5080$ K \citep{Kenyon1995} and $T_{\rm eff}=5945$ K \citep{Calvet2004}. This discrepancy motivated a new analysis of the stellar temperature. We obtained an independent estimate of $T_{\rm eff}$ by comparing the optical spectrum of RY Tau obtained with WHT/UES (see Sect.\,\ref{UES}) to a set of synthetic models of the stellar photosphere computed from the ATLAS and SYNTHE codes \citep{Kurucz1993, Sbordone2004}. We varied $T_{\rm eff}$ and surface gravity log($g$) and found the best fit from the model with \mbox{$T_{\rm eff}=5750$ K} and log($g$)=3.58 (see Appendix \ref{Spec_type} for details). 

We re-calculated all stellar properties using the newly derived $T_{\rm eff}$ and $d$=133 pc (see Sect.\,\ref{RYTau_distance}). We collected the optical-to-millimeter photometry of the source from the literature. RY Tau is known to show a short (weeks) and a long (years) irregular variability in the visible and NIR \citep[see e.g.,][]{Herbst1999}, with variations up to 2.3 mag in the V band. In particular, the brightness is anti-correlated with the linear polarization \citep{Herbst1994, Oudmaijer2001}, suggesting that the short-term variability is due to transient occultations by circumstellar material (e.g., the UX Ori phenomenon). In this scenario, the maximum observed photometry at a given state of the long-term variability is more representative of the stellar photosphere. We set the benchmark photometry to the upper tertile $V$=10.0 mag since this represents the typical high state of the short (weeks) variability during an intermediate state of the long (years) variability. This choice is further motivated by the discussion of Sect.\,\ref{Discussion_episodic}. The correctness of this approach is also supported by the SED, where the adopted optical brightness is in good agreement with the near- and mid-IR photometry which is less subject to variability (see Appendix \ref{Spec_type}).

\begin{table}
\centering
\caption{Stellar properties from this work.}
\label{Table_star}
\begin{tabular}{ll}
\hline
Distance & 133 pc \\
Effective temperature & 5750 K \\
Extinction $A_{\rm V}$ & 1.5 $\pm$ 0.7 mag \\
Luminosity & $6.3^{+9.1}_{-3.2}\ {\rm L_{\odot}}$ \\
Mass & $\approx$1.9 $M_{\odot}$ \\
Age & $\approx4.5$ Myr \\
\hline
\end{tabular}
\end{table}

Then, we calculated the visual extinction $A_{\rm V}$ through the extinction law by \citet{Cardelli1989} with $R_{\rm V}$ = 3.1 and from the comparison of synthetic and observed (V$-$R), (V$-$I), and (R$-$I) colors. Given the high stellar variability, we performed this calculation across the entire light curve by \citet{Herbst1999} and found that the average $A_{\rm V}$ calculated over few tens of measurements decreases from 1.7 to 0.6 with increasing visual brightness (see Fig.\,\ref{Spectral_type}). Here we adopted $A_{\rm V}=1.5\pm0.7$ based on the photometry corresponding to the aforementioned $V$=10.0 mag. The uncertainty derives from the standard deviation of the sub-sample with $V=10.0\pm0.4$ mag. We calculated the stellar luminosity $L_*$ by integrating the flux of the model scaled to the de-reddened photometry and obtained ${L_*=6.3^{+9.1}_{-3.2}\ {\rm L_{\odot}}}$\footnote{This uncertainty is obtained from the uncertainty on $A_{\rm V}$ and on the $\sigma(d)$ of the L1495 cloud sample (5.4 pc, see Sect.\,\ref{RYTau_distance}). By adopting the Gaia DR1 measurement of $d$=176 pc, our calculation yielded ${L_*=11.2\ {\rm L_{\odot}}}$.}. This relatively low luminosity is peculiar of a star slightly older than the typically assumed Taurus age (1$-$2 Myr). In fact, different sets of PMS tracks \citep[Dartmouth, MIST, PARSEC,][]{Dotter2008, Dotter2016, Bressan2012} agree in identifying a stellar mass of 1.9 M$_{\odot}$ at an age of $\approx4.5$ Myr. These estimates should be taken with caution given the large uncertainties highlighted here and in Sect.\,\ref{RYTau_distance}. 

    
\begin{figure*}
  \centering
 \includegraphics[width=18cm]{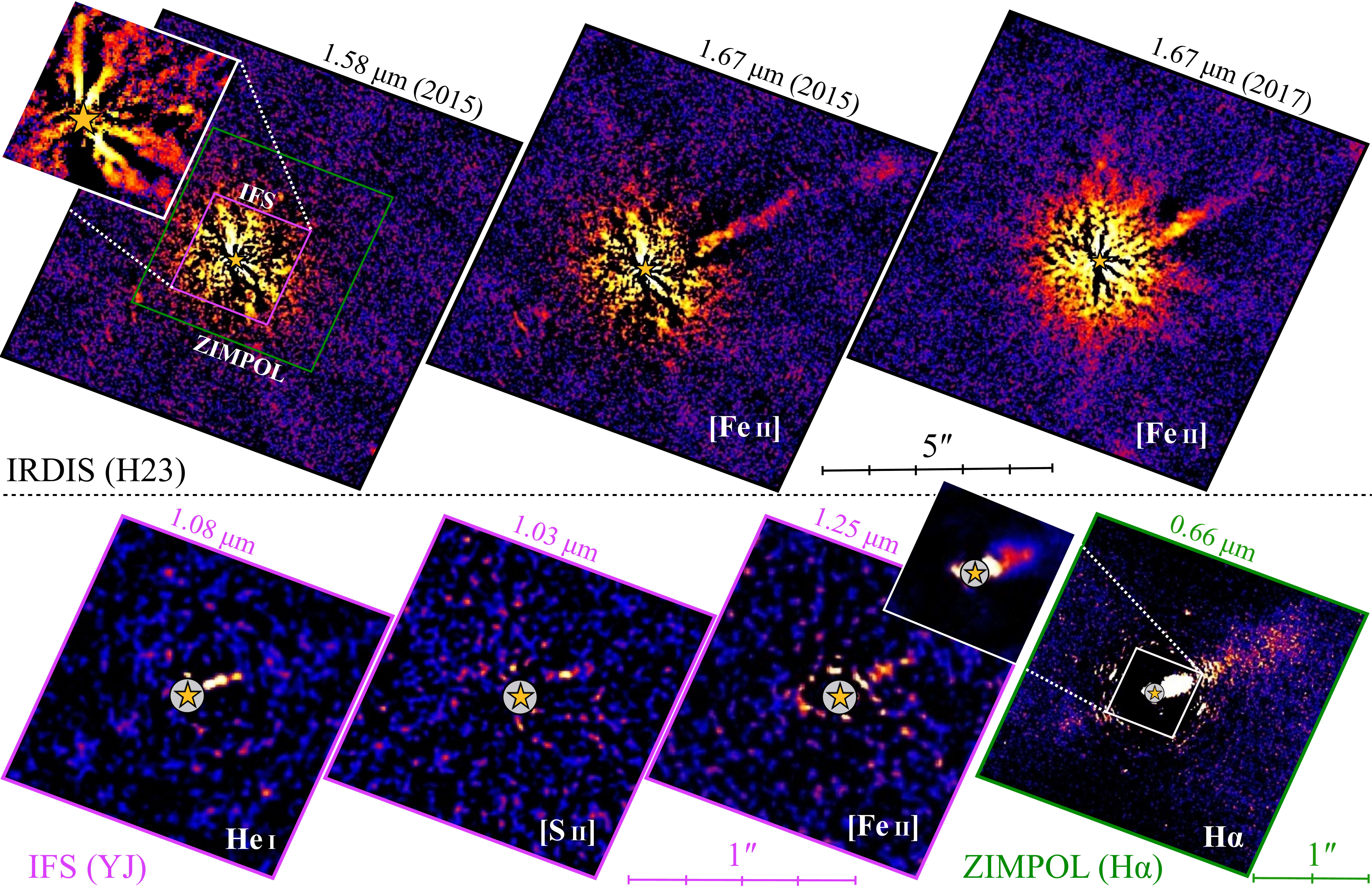} 
      \caption{Imagery of the jet of RY Tau from IRDIS (top), IFS (bottom left), and ZIMPOL (bottom right). The first two IRDIS panels are from 2015, with the left one being in the H2 band and the middle one in the H3 band. The right panel is the H3 image taken in 2017. These images are shown in logarithmic scale with the first two only having the same color stretch. IFS panels are specific channels corresponding to $\lambda$=1.08, 1.03, and 1.25 $\mu$m while the ZIMPOL image is in the H$\alpha$ filter at 0.66 $\mu$m. These images have arbitrary linear color stretch. The inset box of the H$\alpha$ image has a {different} color stretch and a spatial scale corresponding to the IFS images. The {violet} box in the top left panel corresponds to the field of the IFS images while the green box corresponds to the field of the main ZIMPOL image. The stellar position is indicated by the symbol. North is up, East is left.}
          \label{Jet_Imagery}
  \end{figure*}

\section{Results from VLT/SPHERE} \label{RY_Tau}

\subsection{The envelope} \label{ZIMPOL_RYTau}
The polarimetric images of RY Tau obtained with ZIMPOL are shown in Fig.\,\ref{Imagery_ZIMPOL}. The $Q$ and $U$ parameters do not show the quadrilobe pattern typical of centro-symmetric scattering from protoplanetary disks \citep[see e.g.,][]{Quanz2011}. Instead, two regions with diffuse positive and negative values are found to North and South. This evident deviation from centro-symmetric scattering may in principle discourage the use of the $Q_{\phi}$ and $U_{\phi}$ parameters, which are typically adopted under the assumption of this type of scattering. Nonetheless, we show these parameters in Fig.\,\ref{Imagery_ZIMPOL} where it can be seen that ($i$) $Q_{\phi}$ consists of two bright wedges to North and to West and three fainter regions to the SE, two of which with negative values (on average $\approx5$ times lower in absolute value than the bright wedge) and ($ii$) the $U_{\phi}$ image appears as a relatively symmetric alternation of positive and negative regions (with the negative regions being on average $\approx25\%$ brighter in absolute value). They therefore significantly diverge from the ideal morphology to be expected in case of centro-symmetric scattering where ($i$) $Q_{\phi}$ only contains positive values and ($ii$) $U_{\phi}$ only contains noise.

The combination of the Stokes $Q$ and $U$ parameters into $PI$ (see the {left panel of Fig.\,\ref{Imagery_ZIMPOL}}, shown in logarithmic scale) results into a double-wedge structure resembling that seen in the $Q_{\phi}$ image. This morphology is to first order consistent with the fractional polarization $PI/I$ by \citet{Takami2013}, even though a closer inspection reveals that the two wedges from SPHERE are more collimated and their azimuthal distance smaller. From the inner region of the image (obtained from the FastPolarimetry exposures described in Sect.\,\ref{Observations}), we found no sign of radial discontinuity for the two wedges at least down to the angular resolution of these observations ($\approx 0.03\arcsec$). More importantly, we found nowhere signs of any circumstellar disk features (like gaps, arms, or rings) within the diffuse signal as it is recurrently seen in {optical and NIR imaging polarimetry} of this type of stars. {Finally, the disk {emission} from ALMA by \citet{Long2018b} does not have any counterpart in our image (see the bottom right panel of Fig.\,\ref{Imagery_ZIMPOL}). These considerations and the strong deviation from centrosymmetric scattering }suggest to refer to this diffuse {scattered light} as a residual envelope rather than a disk, {in agreement with what was concluded by \citet{Takami2013}} (see Sect.\,\ref{Discussion_nebulosity}).

The total intensity image obtained in RDI as described in Sect.\,\ref{Observations_SPHERE} is also shown in Fig.\,\ref{Imagery_ZIMPOL} (right column). The only signal that we could retrieve from this procedure is visible as bright double lobes in proximity of the star. It is clear that these lobes match the innermost regions of the extended wedges visible in polarized light. The poor quality of the RDI image with respect to the DPI images discourages the calculation of the polarized-to-total light ratio. It can nonetheless be presumed that the DPI images trace to large extent the total scattered light distribution and that the observed double-wedge structure is not peculiar of the polarizing properties.

\subsection{The jet} \label{RYTau_Jet}

\subsubsection{Detection of spectral lines} \label{Jet_detection}
We investigated four sets of images of RY Tau taken with IRDIS in DBI, two taken in the H23 band and two in the K12 band (see Sect.\,\ref{Observations}). All images show multiple arms which are spatially consistent with the diffuse polarized emission from ZIMPOL (see Sect.\,\ref{ZIMPOL_RYTau}). This structure can be seen from Fig.\,\ref{Jet_Imagery} and, in particular, in the inset image of the top left panel which roughly corresponds to the spatial field of Fig.\,\ref{Jet_Imagery}. In addition to this structure, the two available H3 images clearly show evidence of the jet (see the top panels of Fig.\,\ref{Jet_Imagery}). This is visible to NW discontinuously from  approximately 1\arcsec\ out to the detector edge at 6\arcsec. There is no sign of the jet from the H2 images nor from any of the K12 images, suggesting that this emission is the \mbox{[Fe \textsc{ii}]} line at 1.64 $\mu$m, which is in fact sitting within the H3 band wavelength coverage (1.61$-$1.72 $\mu$m). The K1 wavelength interval includes the H$_2$ line at 2.12 $\mu$m but this line is not detected. The morphology of both the inner structure and the jet looks very similar from the different post-processing techniques (see Sect.\,\ref{Observations} and Appendix \ref{Post_proc}). Thus, in this paper we only focus on the IRDIS images processed with cADI.  {We nonetheless caution that the process of ADI is known to affect the actual morphology of azimuthally extended structures (see Sect.\,\ref{Jet_motion} for details).}

The jet is also detected in 5 of the 39 IFS channels in the YJ band (see the bottom panels of Fig.\,\ref{Jet_Imagery}). These channels correspond to $\lambda$=1.08 $\mu$m and 1.03 $\mu$m and to three consecutive channels around $\lambda$=1.25 $\mu$m. Three lines are recurrently detected from jets around these wavelengths: He \textsc{i}, [S \textsc{ii}], and [Fe \textsc{ii}], respectively. {The same lines are detected in the YH band but with a lower intensity because the wavewidth of each channel is larger. Thus, we only focus on the YJ dataset.} The emission in He \textsc{i} is relatively strong and appears as a continuous, narrow feature from the coronagraph edge to $\sim0.3\arcsec$, and putatively further out. The [S \textsc{ii}] line is barely detected as a bright knot coincident with the western end of the He \textsc{i} main emission. The [Fe \textsc{ii}] emission appears faint and more radially diffuse than the other lines. Stacking all other channels yielded no further detectable signal indicating that no other lines of comparable brightness are present in our dataset. All lines are best detected in PCA ADI+SDI (see Sect.\,\ref{Observations}) in the 2017 dataset and in this paper we only focus on these sets of images.   

\begin{figure}
  \centering
 \includegraphics[width=7.5cm]{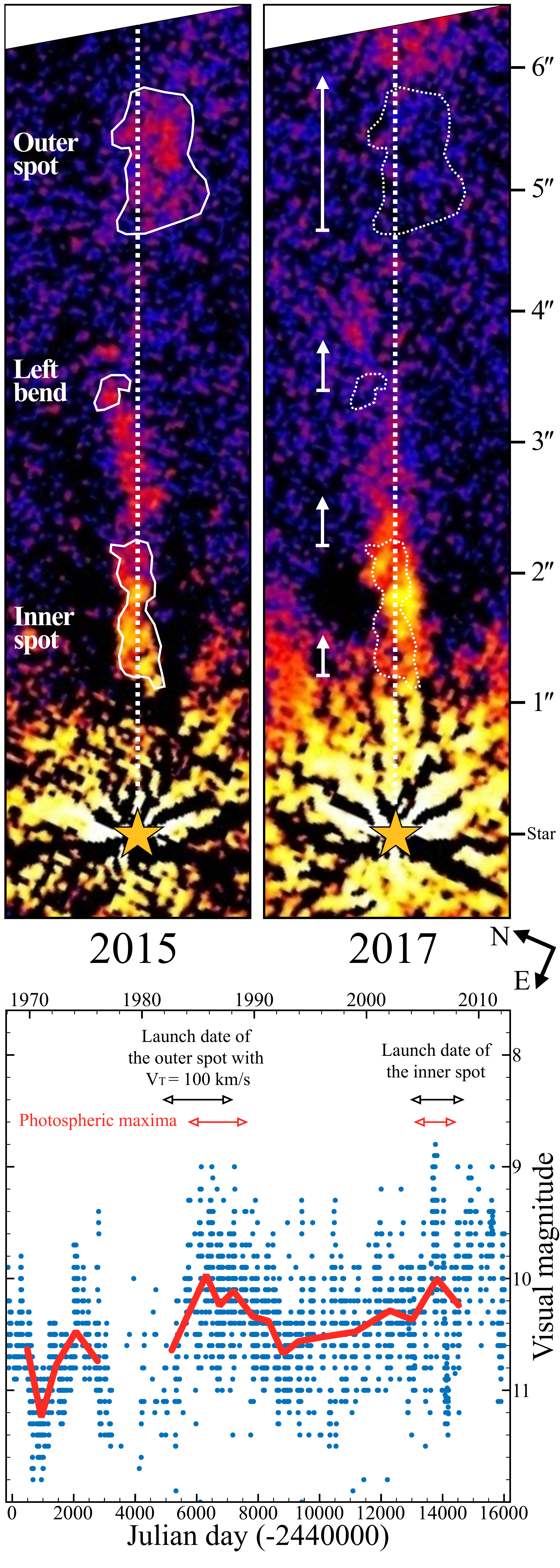} 
      \caption{Jet of RY Tau from the H3 images {tracing the [Fe \textsc{ii}] emission at 1.64 $\mu$m from the jet}. Top panel: comparison of the jet radial extent between 2015 and 2017. {The 2015 features identified in both images are replicated {as dash contours} in the 2017 image. The arrows represent their apparent motion. The jet axis (P.A.=293$\degree$) is indicated by the dashed line. The bright signal in the inner 1\arcsec is primarily stellar flux residual.} Bottom panel: the stellar lightcurve in the last 50 years. The red lines is the best fit to the 3100 blue datapoints. The calculated launch date of the inner and outer jet {spots} are indicated by the black arrows, while the red arrows highlight the local photospheric maxima from the curve.}
          \label{IRDIS_Jet}
  \end{figure}

Finally, we recovered the jet emission from the ZIMPOL/SDI image in the H$\alpha$ filter at 0.66 $\mu$m (see Sect.\,\ref{Observations}). In the inner region, this type of emission appears as a strong narrow feature extending out to $\sim0.3\arcsec$ (see bottom right panel of Fig.\,\ref{Jet_Imagery}) thus resembling the He \textsc{i} main emission. Further out, the signal is much fainter and more azimuthally extended (see Sect.\,\ref{Jet_width}). From this image, we recovered a faint emission from the counter-jet at 180$\degree$ from the main jet.

\subsubsection{Jet proper motion} \label{Jet_motion}
From the IRDIS images, the jet appears discontinuous and marginally curved along its radial extent, as can be seen from Fig.\,\ref{IRDIS_Jet}. Looking at the 2015 dataset, the brightest regions is the inner one (out to 2\arcsec, referred to as inner {spot}), almost no emission is detected from 3.5\arcsec\ to 5\arcsec, but an additional, outer {spot} is evident from 5\arcsec\ to 6\arcsec. The jet looks rather collimated in the inner region (width $\approx0.3\arcsec$) while it appears more azimuthally extended (up to $\approx0.6\arcsec$) further out (see Sect.\,\ref{Jet_width}). A small bend toward the North (left bend) is visible at 3.5\arcsec. Similarly, the outer {spot} appears slightly displaced from the jet axis {\citep[assumed at 293$\degree$ from the disk geometry,][]{Pinilla2018}}. We caution that in principle the process of ADI may bias the morphology of any extended emissions \citep[see e.g.,][]{Garufi2016, Pohl2017}. {This is particularly true for the azimuthal direction since the net effect of the ADI on an azimuthal extended feature is the self-subtraction}. However, this caveat might not be particularly impactful since the RDI image performed on this dataset looks very similar to the ADI images (see Appendix \ref{Post_proc}). Furthermore, the very good angular resolution of these observations ($\approx$0.06\arcsec) is such that all these features are well resolved.

\begin{figure*}
  \centering
 \includegraphics[width=17cm]{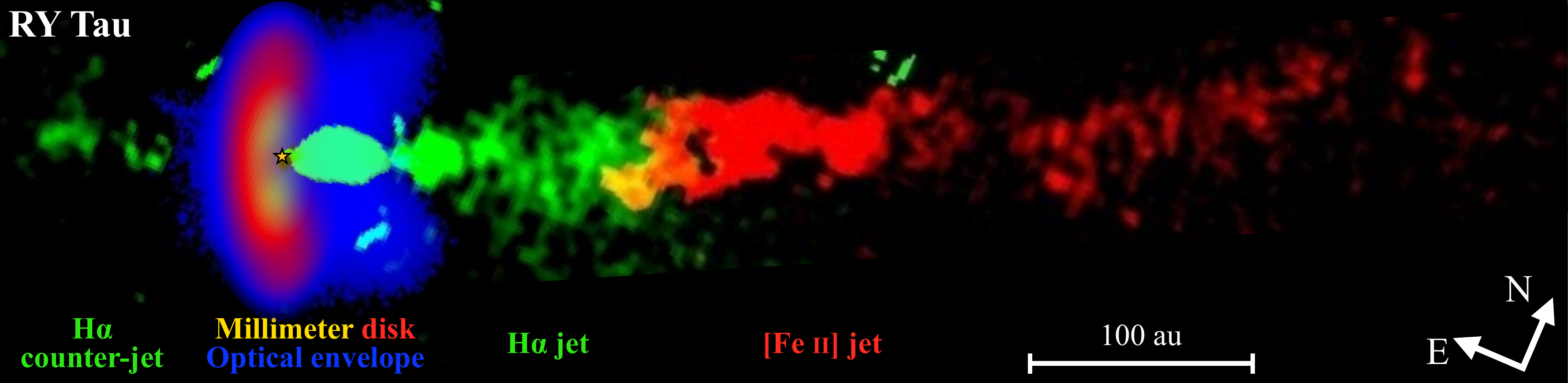} 
      \caption{Composite image of RY Tau comparing the datasets of this work superimposed to the disk seen in the mm by ALMA \citep{Long2018b}. Our data consists of the optical envelope of Fig.\,\ref{Imagery_ZIMPOL} and of the prominent jet traced by H$\alpha$ and [Fe \textsc{ii}] lines of Fig.\,\ref{Jet_Imagery}.}
          \label{Collage}
  \end{figure*}

The 2017 dataset shows the same inner {spot} and, marginally, the left bend. Both structures seem to be offset from the 2015 image (see Fig.\,\ref{IRDIS_Jet}). To quantify this offset, we measured the radial location of the inner and outer edge of the inner {spot} as well as the jet location which is most displaced by the jet axis in the proximity of the left bend. This turned out to be 0.31\arcsec $\pm$ 0.03\arcsec, 0.36\arcsec $\pm$ 0.03\arcsec, and 0.37\arcsec $\pm$ 0.03\arcsec, respectively {(with uncertainties adopted as half of the image resolution)}. It is thus possible that this offset reflects the jet proper motion. Assuming $d=133$ pc (see Sect.\,\ref{RYTau_distance}), these values correspond to a tangential velocity $V_{\rm T} \approx90$, 105, and 110 km s$^{-1}$, which is a reasonable value for a jet but considerably lower than that inferred by \citet[][]{St-Onge2008} from Gemini and Hubble observations further out (165 km s$^{-1}$ at 6\arcsec). The jet radial velocity $V_{\rm R}$ from [O \textsc{i}] emission is 70$-$75 km s$^{-1}$ \citep{Cabrit1990, Coffey2015}, while it is much larger inside $1\arcsec$ {\citep[110$-$150 km s$^{-1}$,][]{Skinner2018}}. The jet inclination deriving from our $V_{\rm T}$ and $V_{\rm R}$=70 km s$^{-1}$ is $i_{\rm jet}=35\degree$ which is slightly larger than that expected (28\degree) from the disk geometry \citep[see][]{Pinilla2018}. Interestingly, the 2017 image does not show the outer {spot} at the location expected from the aforementioned $V_{\rm T}$. Instead, signal with a similar morphology is revealed at the detector edge. Although only the tail of this feature is visible, it is clear from a close inspection of different epochs and post-processing techniques (see Appendix \ref{Post_proc}) that it corresponds to the outer {spot}. If so, the $V_{\rm T}$ of this jet feature is clearly higher than that of the others and we constrained it as $V_{\rm T}\approx 300\ \rm km\, s^{-1}$.

Having estimated $V_{\rm T}$, we calculated the launch date of inner and outer {spots}. Assuming that the $V_{\rm T}$ inferred for the inner {spot} has been constant with time, this feature turned out to be launched between early 2004 and late 2007. To check for {possible} stellar activity level during this period, we retrieved the lightcurve of RY Tau in the visual waveband over the last 50 years from AAVSO\footnote{www.aavso.org}. Interestingly, the possible launch interval of the inner {spot} includes the time of the highest photometric state over this period, between December 2005 and March 2006 (see bottom panel of Fig.\,\ref{IRDIS_Jet}). On the other hand, the launch date of the outer {spot} is more controversial since we cannot conclude whether its very high $V_{\rm T}$ ($\approx 300\ \rm km\, s^{-1}$) has been constant or whether it is due to a recent, and possibly temporary, acceleration (see Sect.\,\ref{Discussion_velocity}). Assuming that this feature has travelled at the same velocity of the inner {spot} ($\approx 100\ \rm km\, s^{-1}$) for most of its lifetime, its launch date corresponds to the other main photospheric maximum registered over the last half a century, in the 80's (see Fig.\,\ref{IRDIS_Jet}). Instead, with the currently measured $V_{\rm T}$ the feature would have a launch date approximately corresponding to that of the inner {spot}. 

In line with such reasoning, we could argue that the faint \mbox{[Fe \textsc{ii}]} emission at 1.25 $\mu$m revealed by IFS in the inner {1\arcsec} is due to the very low stellar state registered over the last five years (with visual magnitude spanning from 10.5 to 11, see Fig.\,\ref{IRDIS_Jet} for reference). It is thus possible that the $1\arcsec-5\arcsec$ region of the jet will look relatively faint over the next decade. The bright emission revealed in H$\alpha$ and He \textsc{i} in the inner region is not necessarily in contrast with this scenario since these lines may be connected to strong stationary shocks in proximity of the star and, as such, not trace the actual volumes of expanding gas.

\subsubsection{Jet width and wiggle} \label{Jet_width}
\begin{figure}
  \centering
 \includegraphics[width=9cm]{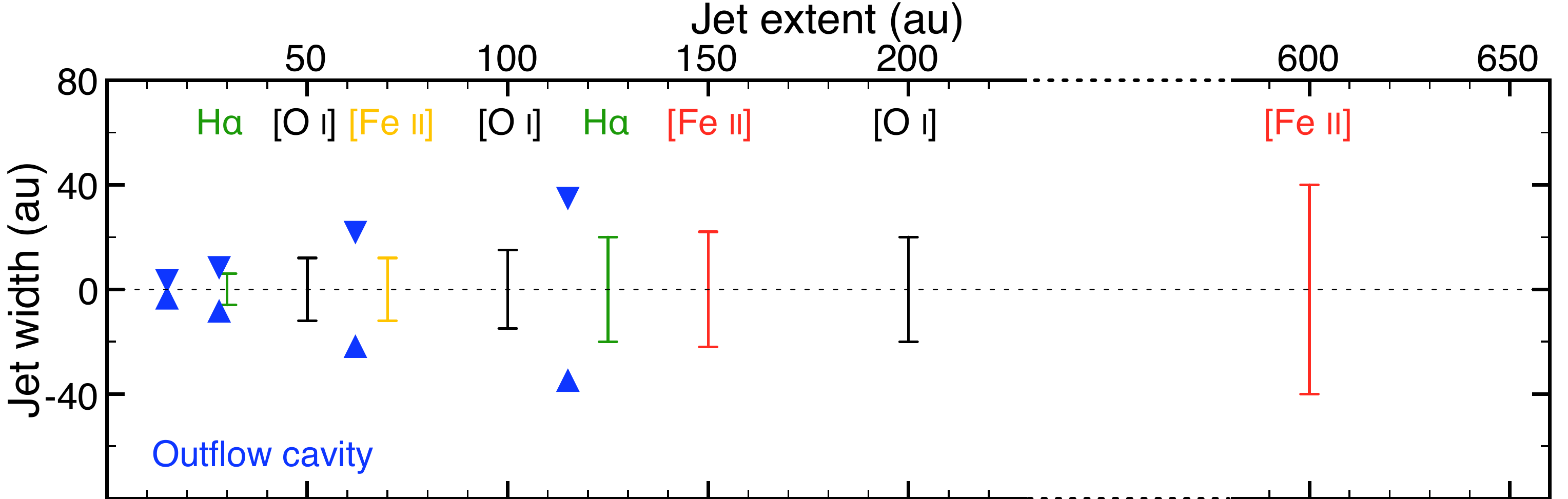}
  \includegraphics[width=9cm]{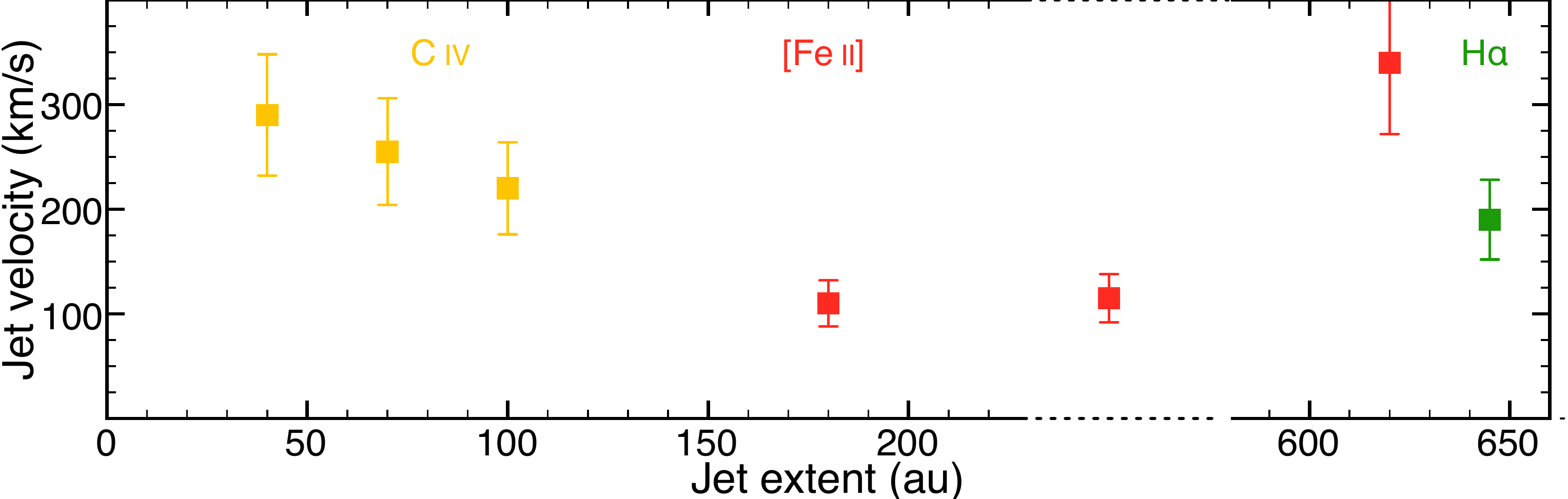}
      \caption{Jet properties with its radial extent. Top: jet width inferred from different tracers from this work, \citet{Agra-Amboage2009} (black), and \citet{Coffey2015} (yellow). The blue triangles indicate the outflow cavity size inferred from {the PI images in Fig.\,\ref{Imagery_ZIMPOL}}. Bottom: jet velocity inferred from different tracers from this work, \citet{Skinner2018} (yellow), and \citet{St-Onge2008} (green). The x-axis has a discontinuity.}
          \label{Jet_width_velocity}
  \end{figure}

The H$\alpha$ and the He \textsc{i} emission from ZIMPOL and IFS appears bright and compact in the inner 40 au. In this region, the jet width measured from these two lines is approximately 12 au and 8 au, respectively. The former estimate is more reliable since the ZIMPOL images have a higher angular resolution ($\sim$4 au vs $\sim$6 au) and were not post-processed with ADI. While we do not recover significant He \textsc{i} emission at larger radii, the H$\alpha$ emission is detected up to 100$-$150 au where the jet looks much wider, i.e.\,$\approx$ 40 au. The inner {spot} detected in \mbox{[Fe \textsc{ii}]} at 200$-$250 au has a comparable width, whereas the outer {spot} at 600 au appears as wide as $\approx$ 80 au. These numbers point to a bottle-neck shape for the jet with an appreciable flow widening occurring between 40 au and 100 au. Interestingly, these radii correspond to the region of increasing azimuthal divergence between the two DPI wedges (see Fig.\,\ref{Imagery_ZIMPOL}). This complementarity can also be appreciated from the diagram of Fig.\,\ref{Jet_width_velocity} and the composite image of Fig.\,\ref{Collage} and it highlights the jet efficiency in sculpting the surrounding medium with its passage.

As commented in Sect.\,\ref{Jet_motion}, the jet orientation changes with the distance from the star. By extracting the center of the azimuthal extension from different tracers and at different radii, the jet position angle P.A.\ varies from 290$\degree$ (from the innermost region and the outer {spot}) to 295$\degree$ (the left bend). Thus, the jet seems to wiggle around the P.A.\ expected from the disk orientation \citep[293$\degree$,][]{Pinilla2018} and the projected half-opening angle of this wiggle is $2\degree-3\degree$. Since the entire opening angle is found between the left bend and the outer knot (see Fig.\,\ref{IRDIS_Jet}), we can infer that a wiggle half-cycle elapses $\approx$300 au corresponding to 15 years (with $V_{\rm T} = 100\ \rm km\, s^{-1}$). If the jet wiggle is also present along the radial direction, this effect may partly reconcile the discrepancy between $V_{\rm T}$ and $V_{\rm R}$ highlighted in Sect.\,\ref{Jet_motion}. 

\section{Discussion} \label{Discussion}

\subsection{The disk and the envelope} \label{Discussion_nebulosity}
Our observations in polarized light of Sect.\,\ref{ZIMPOL_RYTau} are indicative of a partly embedded source. In fact, although the stellar photosphere is visible and the SED is characteristic of Class II objects \citep{Lada1987}, the disk revealed in the sub-mm at \mbox{P.A.\ $=23\degree$} (see Fig.\,\ref{Collage}) remains undetected in scattered light. In the context of {optical and NIR imaging polarimetry} of protoplanetary disks, this morphology is uncommon \citep[see][]{Garufi2018}. Similar observations were described by \citet{Takami2013} who reproduced the polarized emission by means of an optically thin and geometrically thick upper layer. This evident nebulosity is likely acting to scatter the NIR photons from the disk but it is not sufficiently thick to mask the stellar photosphere. In other words, it represents the last stage of a protostellar envelope.

The spatial analogy between {the wedges in the PI image in Fig.\,\ref{Imagery_ZIMPOL}} and the jet of Fig.\,\ref{Jet_Imagery} is reminiscent of the outflow cavity of earlier evolutionary stages \citep[see e.g.,][]{Arce2006}. In particular, the opening angle measured from the azimuthal distance of our DPI wedges ($\approx 80\degree$, {considering the edge as the angle where the flux drops below 3$\sigma$}) is what is expected from $\gtrsim1$ Myr-old sources \citep{Seale2008} where most of the envelope has been swept away by the jet and a broader wind outflow. This effect can be seen from Fig.\,\ref{Jet_width_velocity} and Fig.\,\ref{Collage}, where it is clear that the jet appears less collimated from 50 au outward, that is where the opening angle between the optical wedges appreciably increases.

It is intriguing that in this scenario of early evolution, the protoplanetary disk already shows an inner dusty cavity \citep{Isella2010}. According to recent ALMA data (see Fig.\,\ref{Collage}), this cavity is relatively large ($\sim$18 au) but also rather shallow having only a factor 2 depletion in dust \citep{Long2018b}. The existence of both a disk cavity and a jet is not necessarily odd provided that gas is still flowing within the cavity. The source is in fact accreting \citep[$\approx 2.2\cdot10^{-8}\,{\rm M_{\odot}/yr}$,][]{Mendigutia2011}. It is tempting to speculate that the cavity has formed recently and that this has not impacted yet on the jet launch mechanisms (see also Sect.\,\ref{Discussion_sample}).

\subsection{The jet} \label{Discussion_jet}
As discussed in Sect.\,\ref{RYTau_Jet}, we detected the jet in the H$\alpha$, [S \textsc{ii}] 1.03 $\mu$m, He \textsc{i} 1.08 $\mu$m, and [Fe \textsc{ii}] lines at 1.25 $\mu$m and 1.64 $\mu$m. Previous work on RY Tau has revealed the emission of the H$\alpha$ and other optical lines \citep{Cabrit1990, Hartigan1995, St-Onge2008, Agra-Amboage2009} and of the [Fe \textsc{ii}] in the NIR \citep{Coffey2015}, whereas the [S \textsc{ii}] and the He \textsc{i} are detected for the first time {in this jet.} 

{Such a variety of detected emission lines corresponds to the presence of gas in widely different conditions of ionization and excitation level. This is consistent with the presence of multiple shocks in the beam, as it is frequently observed in the initial section of stellar jets \citep[see e.g., the sub-arcsecond maps from DG Tau or HH 30 star][]{Bacciotti2000, Hartigan2007}. Such shock fronts can produce widely different excitation/ionization conditions because of a combination of effects. In fact, each front can have a different propagation velocity because of different ejection velocity at the jet base, and subsequent merging with the adjacent fronts.} 

{Furthermore, the bow-like shape of most of these fronts produces variations of the shock velocity (which drives the excitation level), as the latter is a function of the local inclination angle of the front with respect to the jet direction \citep[e.g.,][]{Hartigan1987}. Finally, even for a plane-parallel shock, the cooling of the gas behind the front produces strong gradients in the gas density, temperature and ionization which  give rise to lines of very different excitation in a spatially unresolved layer \citep[see e.g.,][]{Hartigan1994, Nisini2005}. We note however, that the He \textsc{i} at 1.08 $\mu$m has a high excitation energy ($\sim$ 20 eV), which means that shock velocities larger than 100 km s$^{-1}$ must be present in the inner 0.5\arcsec\ of the jet if this line is collisionally excited \citep[see tables in][and Sect.\,\ref{Discussion_collimation}]{Hartigan1987}}. 

Our analysis on the jet of RY Tau revealed that:

\begin{enumerate}

\item The activity stage of RY Tau might have a temporal connection with the launch of the jet {spots}.

\item The jet is very collimated ($\sim$12 au width) on spatial scales of less than 40 au {from the star}.

\item The inner part ($\lesssim$500 au) of the jet is currently moving as a rigid body with $V_{\rm T}\approx100$ km s$^{-1}$. However, the outer {spot} at $\sim$650 au appears to travel three times faster.

\item The jet is wiggling in 30 years with an opening angle of $\approx5\degree$.

\end{enumerate}

\subsubsection{Episodic accretion and ejection} \label{Discussion_episodic}
Young stars are known to undergo stages of episodic accretion. The physical origin of these episodes is highly debated but it is likely related to the intrinsic inability for the inner disk to accrete at a steady state, to density perturbations within the disk, or to variability of the magnetic field \citep[see the review by][]{Armitage2015}. In magneto-centrifugal models, accretion and ejection activities are predicted to be coupled \citep[see the review by][]{Konigl2000}, and the jet bears record of these episodic {accretion}.

In view of this, the likely temporal analogy between the jet {spot} launching and the state of RY Tau points toward an increased mass outflow rate that would in turn be connected to an episodic accretion occurred in 1985 and 2006 (see Fig.\,\ref{IRDIS_Jet}). A similar behavior was shown in another well known jet source, HD163296. Unlike RY Tau, this Herbig star has been photometrically stable over many decades. Yet, a few photometric dimming events \citep{Ellerbroek2014} have been ascribed to the episodic obscuration from circumstellar dusty clouds while optical/NIR brightening events \citep{Sitko2008} to the periodic lifting of dusty material into disk wind or halo \citep{Vinkovic2006, Bans2012}. 

\begin{figure}
  \centering
 \includegraphics[width=9cm]{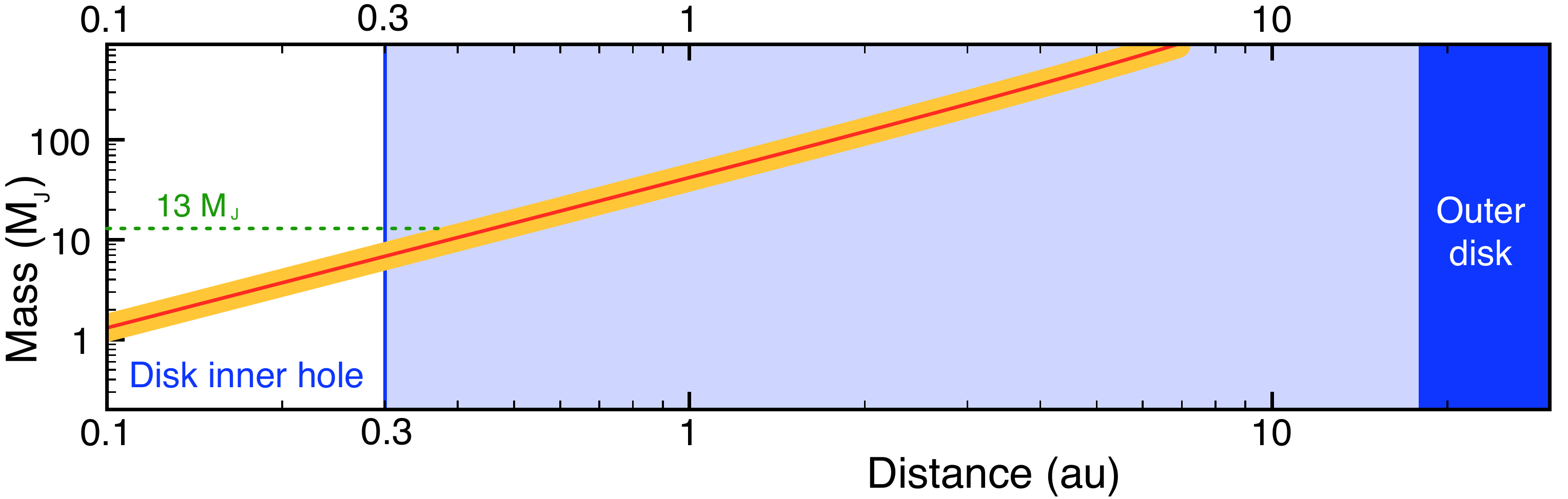} 
      \caption{Mass and orbital distance of a putative companion responsible for the jet wiggle in RY Tau. The yellow region indicates the analytical error bars. The cyan region is between the inner edge of the outer disk from ALMA data (18 au) and the inner edge of the inner disk from mid-IR interferometric data (0.3 au). The green dashed line is the formal threshold between giant planet and brown dwarf.}
          \label{Companion}
  \end{figure}

Both these effects are more erratic in RY Tau \citep[see][]{Petrov1999} but the physics responsible for this behavior is likely the same. The younger age of RY Tau ($\approx$5 vs 10 Myr), the apparently earlier stage of the circumstellar material morphology, and their similar stellar mass may suggest that RY Tau and HD163296 represent different stages of a similar evolutionary path.

\subsubsection{Jet collimation and recollimation shock} \label{Discussion_collimation}
The unprecedented resolution of these images ($\sim$30 mas for ZIMPOL and 40$-$60 mas for IRDIS/IFS) and the detection of multiple jet tracers allow an accurate measurement of the jet width from 40 au to 600 au. The jet opening angle derived from the estimates of Sect.\,\ref{Jet_width} is $\approx3.5\degree$. This opening angle is consistent with the widening due to a highly supersonic jet propagating with an average velocity of 150 km s$^{-1}$ and with an internal sound speed of about $\sim$10 km s$^{-1}$ (assuming a typical gas temperature of $10{^4}$ K). This indicates that the jet collimation occurs on spatial scales smaller than 50 au as predicted by magneto-centrifugal models \citep[e.g.,][]{Pudritz2007} and observed in a few other jets from TTSs imaged by HST with 0.1\arcsec-resolution \citep[see e.g.,][]{Cabrit2007}. In all cases, including RY Tau, the measured widths are consistent with those imposed by the collimation via magnetic hoop stresses in models of disk winds, as illustrated for example in Fig.\,2 of \citet{Ray2007}. The magnetically collimated jet excavates the surrounding medium in its passage, as it results from the complementary morphology of the outflow cavity shown in Sect.\,\ref{Jet_width}.

Interestingly, the innermost {50 au from the star} where the jet is already collimated shows a clear emission in the He \textsc{i} line at 1.08 $\mu$m. Given its high excitation energy ($\sim$20 eV), the He \textsc{i} emission can be produced either collisionally (if the gas is heated up to $2.5\cdot10^4$ K) or by recombination (if the gas is highly ionized). In turn, these scenarios require a high velocity shock and/or a source of ionizing radiation. This is consistent with the detection of faint X-ray emission extending a few arcseconds along the jet since this probes hot plasma \citep[$T\sim10^6$ K,][]{Skinner2011}, and of other high-excitation lines detected at UV wavelengths close to the driving source, like the C \textsc{iv} doublet at 1548/1550 $\AA$ \citep[$T\sim10^5$ K,][]{Skinner2018}, and the C \textsc{iii} 1908 $\AA$ and S \textsc{iii} 1892 $\AA$ lines \citep[$T\sim3\cdot10^4$ K,][]{GomezdeCastro2001}.

The detection of the He \textsc{i} line at 1.08 $\mu$m and of other high-excitation lines might also be the signature of the so-called recollimation shock that forms when an initially wide wind is collimated into a jet, likely by magnetic fields \citep[see e.g.,][]{Bonito2011, Frank2014}. Previous detections of the He \textsc{i} line at 1.08 $\mu$m have been reported around a large sample of TTSs \citep{Edwards2006} {but the emission was unresolved. However, for DG Tau it was shown through a spectro-astrometric analysis that the He \textsc{i} emission was elongated along the P.A. of the blue-shifted outflow lobe, and it was kinematically associated with the resolved region bright in [Fe \textsc{ii}], which traces the collimated jet \citep{Takami2002}. In the case of our AO observations of RY Tau, He \textsc{i} emission is spatially resolved, it is clearly elongated and is likely spatially coincident with the first $\sim$50 au of the collimated jet seen in H$\alpha$ and in the other forbidden lines. This confirms that the energetic shocks giving rise to the He \textsc{i} emission are related to the outflow activity.}
  
\subsubsection{Jet velocity} \label{Discussion_velocity}
The velocity inferred in Sect.\,\ref{Jet_motion} (a de-projected $\approx105-125$ km s$^{-1}$) is quite moderate for a jet \citep[see][]{Frank2014}. \citet{Skinner2018} noticed that the jet velocity of RY Tau rapidly decreases from 300 km s$^{-1}$ at 40 au to 240 km s$^{-1}$ at 100 au. Interpolating their trend yields the velocity that we inferred for the region outside 150 au (see Fig.\,\ref{Jet_width_velocity}). Furthermore, the inner {spot} and the left bend of Fig.\,\ref{IRDIS_Jet} can be presumably recognized in the GEMINI-NIFS image taken by \citet{Coffey2015} from the same tracer in 2009. The observed radial offset results in an intrinsic velocity of $\approx125$ km s$^{-1}$ which is consistent with our measurement. Interestingly, the spatial interval of the rapid jet braking (from 40 to 150 au) corresponds to the abrupt increase of jet width (see Sect.\,\ref{Discussion_collimation} and Fig.\,\ref{Jet_width_velocity}), suggesting that the jet velocity and width are connected.  

The very different velocity of the outer {spot} ($\approx$340 km s$^{-1}$) with respect to the inner jet is controversial. It could be possibly associated to a bright knot at 1.3\arcsec\ detected in the [O \textsc{i}] by \citet{Agra-Amboage2009} in 2002. Its average velocity from 2002 to 2015 would be $\approx$215 km s$^{-1}$, that is an intermediate value between the estimated velocities of inner and outer jet. In principle, an abrupt decrease of the ambient density acts to accelerate the jet. It can be calculated from Eq.\,(5) of \citet{Raga1998} that a 3 times faster jet requires a $\sim15-50$ times less dense medium (in $\sim$150 au). The jet velocity obtained slightly outside of our outer {spot} by \citet[190 km s$^{-1}$]{St-Onge2008} instead only requires a factor 5$-$10 of density decrease (in $\gtrsim300$ au).

\subsubsection{Disk warp}
The jet wiggle inferred in Sect.\,\ref{Jet_width} can have two major origins, namely ($i$) the presence of a stellar companion massive enough to induce an orbital motion for the primary star \citep{Anglada2007} and ($ii$) the precession of an inner disk portion with respect to the outer disk \citep{Zhu2019}.

In the former case, the launching area of the jet moves as the primary star orbits the center of mass. The secondary star must therefore be rather massive. We employed Eq.\,(7) and (9) of \citet{Anglada2007} to infer the mass of a putative companion necessary to induce the observed jet wiggle. Using our observed period of the wiggle, half-opening angle (see Sect.\,\ref{Jet_width}), and jet velocity (see Sect.\,\ref{Jet_motion}), the calculation yielded an object of mass $M_{\rm c}=1.1\,{\rm M_{\odot}}$ at a separation of 14 au from RY Tau {(corresponding to $\gtrsim0.1\arcsec$)}. However, such an object would be easily recovered from our intensity images. In general, although the presence of a stellar companion has been claimed by several authors \citep[e.g.,][]{Bertout1999, Nguyen2012} this has not been firmly demonstrated, and RY Tau is still {considered} as a single star by recent {mid-IR} interferometric campaigns \citep[e.g.,][]{Varga2018}.  

All this said, the precession of a disk warp or broken inner disk \citep[e.g.,][]{Min2017, Facchini2018} seems a more likely explanation for the jet wiggle. This scenario may also require the presence of a companion, although it does not need to be so massive to alter the orbital motion of RY Tau. As an alternative to a companion, a disk warp may also be induced by the disk misalignment with the stellar magnetic axis \citep[see e.g.,][]{Bouvier1999} but in this scenario the precessing timescale is related to the stellar rotation and is thus too rapid to explain our jet wiggle. Instead, we focused on the companion scenario and constrained the mass $M_{\rm c}$ and orbital radius $R_{\rm c}$ of this putative companion. We assumed that the jet wiggle is a direct consequence of a precessing disk warp and employed the analytical Eq.\,(27) by \citet{Zhu2019}, that can be re-written as:  
\begin{equation}
T_{\rm pre}=\frac{8}{3q\cos (i)}(1+q)^{1/2}\left(\frac{GM_*}{R_{\rm c}^3}\right)^{-1/2}\left(\frac{R_{\rm c}}{R_{\rm d}}\right)^{3/2}
\end{equation}
where $T_{\rm pre}$ is the precession timescale, $M_*$ is the mass of RY Tau, $q$ is $M_{\rm c}/M_*$, $i$ is the companion orbital/disk angular momentum vector misalignment, and $R_{\rm d}$ the location of the gap induced by the companion. Based on the properties of the jet wiggle, we imposed $i=5\degree$, $T_{\rm pre}=30$ years and assumed that $R_{\rm c}\gtrsim R_{\rm d}$. This left us with the family of solution shown in Fig.\,\ref{Companion}. It turned out that a giant planet inside the disk inner hole \citep[0.3 au,][]{Schegerer2008} or a sub-stellar companion further out could account for the observed jet wiggle. In the latter case, the inferred companion would be massive enough to {tidally} break the inner and outer disk \citep[e.g.,][]{Pinilla2012} but may be small enough to remain undetected in high-contrast imaging and interferometric campaigns. This is particularly true in the inner 2$-$3 au where our calculation yielded $M_{\rm c}< 0.1\, \rm M_{\odot}$. 

\begin{figure}
  \centering
 \includegraphics[width=9cm]{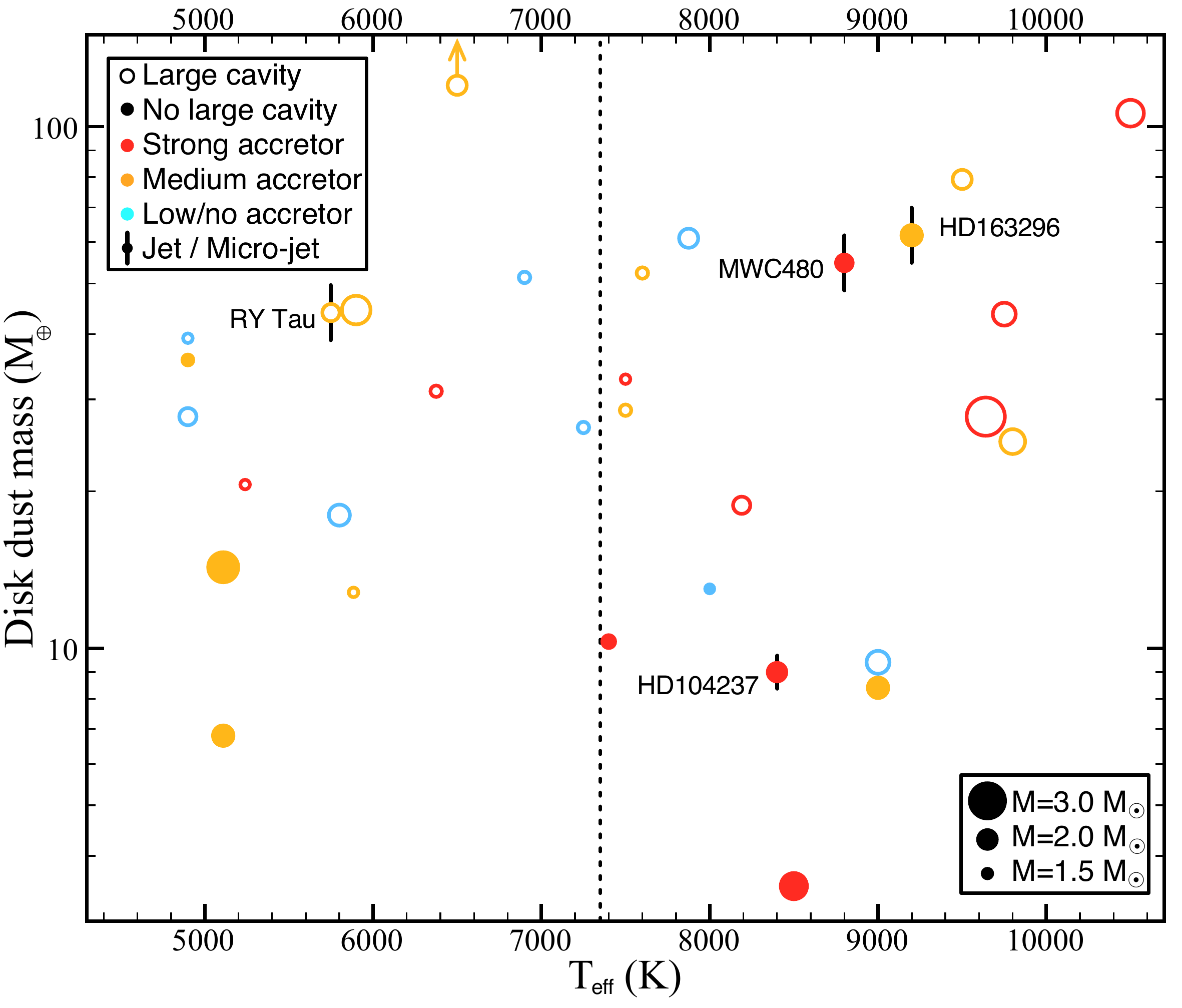} 
      \caption{Disk dust mass vs effective temperature for a representative sample of intermediate-mass stars \citep[see][]{Garufi2018}. Empty symbols indicate the presence of a disk cavity while the large/small horizontal bars the presence of a known jet/micro-jet. Symbol sizes are proportional to the stellar mass with dynamic range from 1.5 $\rm M_{\odot}$ to 3 $\rm M_{\odot}$ while the accretion rate is indicated by the color with strong accretors ($>4\cdot10^{-8}$ M$_{\odot}$/yr) in red, medium accretors (down to $10^{-8}$ M$_{\odot}$/yr) in orange, and low/no accretors in cyan. The dashed line separates TTSs and Herbig stars according to the formal threshold at F0 type.}
          \label{Sample_Jet}
  \end{figure}

\subsection{Jets of other intermediate-mass stars} \label{Discussion_sample}
The occurrence of a known jet around Class II intermediate-mass stars is relatively low {\citep[see e.g.,][]{Corcoran1997, Eisloffel2000}}. In this regard, RY Tau is a fairly exceptional object. In fact, the only other such objects within 200 pc with a direct detection of jet-tracing optical/NIR forbidden lines are, to our knowledge, HD163296 \citep{Devine2000}, MWC480 \citep{Grady2010}, and HD104237 \citep{Grady2004}. Although these four stars are fairly similar in mass ($\sim$2 M$_{\odot}$), their structure is not necessarily identical. Unlike the others, RY Tau is in fact a TTS and as such still possesses a convective envelope that sustains an axisymmetric magnetic field with strong dipole components \citep{Gregory2012}.

In Fig.\,\ref{Sample_Jet}, we compare RY Tau with the sample of nearby intermediate-mass stars with resolved NIR observations of the disk \citep[selected from][with the addition of HD104237]{Garufi2018}. In this sample, albeit not complete in any physical sense, only 12\% of the source is known to host a jet. We must nonetheless caution that the detection of a jet in intermediate-mass stars is intrinsically hampered by the bright central continuum emission. Yet, some considerations on the diagram can be made. First of all, those of HD163296 and MWC480 are by far the most massive full disks {(i.e., hosting no central cavity)} in the sample\footnote{Contrary to the literature, HD142666 is treated as a disk with a cavity, based on \citet{Rubinstein2018}.}. Formally speaking, the disk of RY Tau is not a full disk \citep{Long2018b} but it clearly differs from the other transition disks of this sample. In fact, the SED does not show evidence of a large cavity and the source is still clearly embedded in the natal cloud (see Sect.\,\ref{Discussion_nebulosity}). On the other hand, the possible connection between the disk mass and the presence of a jet is not supported by HD104237, being its disk in the low tail of dust mass distribution. Figure \ref{Sample_Jet} also shows that there is no clear connection between the presence of the jet and the accretion rate nor the stellar mass. In particular, the accretion rate of the four jet-sources is in fact relatively high, {spanning from $3\cdot10^{-8}$ M$_{\odot}$/yr to $2\cdot10^{-7}$ M$_{\odot}$/yr in a sample with an average of  $4\cdot10^{-8}$ M$_{\odot}$/yr (that does not include 3 out of 32 non-detections). However,} other stronger accretors are not known to possess a jet. 

RY Tau is the only IMTTS (over 14) showing a jet while 3 (out of 18) Herbig AeBe are known to host a jet. Bearing in mind the very low statistics, this trend does not support a scenario where jets originate from the stellar surface \citep{Sauty1994}, unless the magnetic fields of Herbig stars are much stronger than what is currently believed \citep[e.g.,][]{Hubrig2018}. On the other hand, a disk-wind scenario where the jet is launched by the magnetic field threading the disk \citep[e.g.,][]{Blandford1982, Konigl2000} does not require a highly magnetized star. Instead, the presence of a full disk is a condition that intuitively favors this mechanism since the launching region of an atomic jet is the inner 4$-$5 au \citep{Bacciotti2002, Coffey2004}. Furthermore, the presence of prominent jets in RY Tau, MWC480, and HD163296 could indicate that more massive disks are more efficient at launching powerful jets whereas less massive disks \citep[that can be as small as 10 au, see][]{Garufi2017} may only launch a micro-jet \citep[like HD104237,][]{Grady2004} or not launch any jet at all. 

The absence of other known nearby IMTTS with a jet could also be a selection bias. In fact, \citet{Garufi2018} noticed a desert in the distribution of imaged IMTTS younger than 4$-$5 Myr and argued that this could be due to the embedded nature of these sources making them a challenging target for AO-driven observations. The intermediate nature of RY Tau discussed in Sect.\,\ref{Discussion_nebulosity} is consistent with this scenario. Stars with $M\sim2-3\,{\rm M_{\odot}}$ may be concealed for several Myr and then rapidly disperse the envelope and lose the jet thus as to appear like the recurrently imaged intermediate-mass stars older than $\sim$5 Myr. 

Finally, RY Tau lies in a rather dense environment, being one of the few intermediate-mass stars in the Taurus cloud. This is in contrast with the majority of the evolved intermediate-mass stars observed thus far and could speculatively mean that the ISM accretes onto the protoplanetary disk for a longer timescale than what occurs in the isolated Herbig stars. This scenario would reconcile the apparent discrepancy between the early stage of the circumstellar environment and the relatively old age of RY Tau, and thus explain how is the jet sustained for a long timescale.

\section{Summary and conclusions} \label{Conclusions}
The intermediate-mass ($M\sim1.9\,{\rm M_{\odot}}$) star RY Tau has been studied with VLT/SPHERE optical and NIR images. In contrast to the vast majority of the PMS stars available from the literature, our scattered-light images do not reveal the protoplanetary disk that is visible at millimeter wavelengths. Instead, we only resolved the inner region of the remnant envelope, {as for previous NIR observations}. This nebulosity shows a characteristic dip along the jet axis that is consistent with the late stages of evolution of outflow cavities.

We detected the atomic jet (and marginally the counter-jet) with an unprecedented intrinsic spatial resolution (down to 4 au) in the H$\alpha$, [S \textsc{ii}] at 1.03 $\mu$m, He \textsc{i} at 1.08 $\mu$m, and [Fe \textsc{ii}] lines in the 1.25 and 1.64 $\mu$m. Having two observing epochs, we constrained from the proper motion a jet tangential velocity of approximately 100 km s$^{-1}$. From this, we inferred a launching date for the main jet {spot} occurred around 2006, when the star experienced a main photometric maximum. The other main stellar brightening of RY Tau occurred in 1985 and can be associated to the launch of an outer {spot}. This finding is a {possible} evidence of the episodic nature of the accretion process and of the intimate connection between accretion and ejection events.

The jet shows a wiggling, bottle-neck shape with a flow widening of a few degrees between 40 and 100 au. The jet morphology in the inner 200 au is consistent with early strong collimation in a magnetic disk wind, supporting once again the magneto-centrifugal scenario as the origin of jets. We also constrained the mass and orbital distance of a putative companion responsible for the observed jet wiggle. A giant planet at sub-au scales or a sub-stellar companion further out may in fact induce a disk warp or a misalignment of the inner disk. Such an object may have escaped all planet-search campaigns carried out thus far.

Finally, we put RY Tau in the context of the most studied intermediate-mass stars and noticed that three of the four known jet-sources, including RY Tau, host a very massive disk. In particular, HD163296 and MWC480 host the most massive full disks of the analyzed sample while the disk of RY Tau only shows a very shallow (possibly recent) cavity that does not leave any imprint on the SED. While it is still possible that the paucity of other IMTTSs with a jet is an observational bias, our small sample of intermediate-mass stars suggests a marginal role for the stellar magnetic field and that more massive disks are more efficient at launching powerful jets. In any case, a deep characterization and a more in-depth statistical analysis of jet-driving sources will be necessary to better constrain the connection between the natal envelope, the protoplanetary disk, and the jet.  

This work also serves as a showcase of the SPHERE capabilities in the search and characterization of the optical jets, following previous work on more distant objects \citep{Antoniucci2016, Schmid2017}. SPHERE allows us to directly image the inner few hundreds au of the jet with a few-au-resolution, in a framework where only spectral observations or coronagraphic imaging at larger distance have been conducted thus far. 

\begin{acknowledgements}
     We are grateful to Deirdre Coffey, Stefano Facchini, Giuseppe Lodato, Elena Pancino, and Paola Pinilla for the useful discussions, to Feng Long for sharing the ALMA data of RY Tau, and to Benjamin Montesinos for the help with the WHT/UES spectra. {We thank the referee, Michihiro Takami, for the insightful comments that improved the manuscript.} We also thank the ESO technical operators at the Paranal observatory for their valuable help during the observations. This work has been supported by the project PRIN-INAF 2016 The Cradle of Life - GENESIS-SKA (General Conditions in Early Planetary Systems for the rise of life with SKA). We also acknowledge support from INAF/Frontiera (Fostering high ResolutiON Technology and Innovation for Exoplanets and Research in Astrophysics) through the "Progetti Premiali" funding scheme of the Italian Ministry of Education, University, and Research. FMe acknowledges funding from ANR of France under contract number ANR-16-CE31-0013. CPe acknowledges financial support from the ICM (Iniciativa Cient\'ifica Milenio) via the N\'ucleo Milenio de Formaci\'on Planetaria grant, from the Universidad de Valpara\'iso. AZu acknowledges support from the CONICYT + PAI/ Convocatoria nacional subvenci\'on a la instalaci\'on en la academia, convocatoria 2017 + Folio PAI77170087. SPHERE is an instrument designed and built by a consortium consisting of IPAG (Grenoble, France), MPIA (Heidelberg, Germany), LAM (Marseille, France), LESIA (Paris, France), Laboratoire Lagrange (Nice, France), INAF - Osservatorio di Padova (Italy), Observatoire de Geneve (Switzerland), ETH Zurich (Switzerland), NOVA (Netherlands), ONERA (France), and ASTRON (Netherlands) in collaboration with ESO. SPHERE was funded by ESO, with additional contributions from the CNRS (France), MPIA (Germany), INAF (Italy), FINES (Switzerland) and NOVA (Netherlands). SPHERE also received funding from the European Commission Sixth and Seventh Framework Programs as part of the Optical Infrared Coordination Network for Astronomy (OPTICON) under grant number RII3-Ct-2004-001566 for FP6 (2004-2008), grant number 226604 for FP7 (2009-2012), and grant number 312430 for FP7 (2013-2016). This work has made use of the SPHERE Data Centre, jointly operated by OSUG/IPAG (Grenoble), PYTHEAS/LAM/CeSAM (Marseille), OCA/Lagrange (Nice) and Observatoire de Paris/LESIA (Paris) and supported by a grant from Labex OSUG@2020 (Investissements d’avenir – ANR10 LABX56). It has also made use of the SIMBAD database, operated at the CDS, Strasbourg, France. 
     \end{acknowledgements}

\bibliographystyle{aa} 
\bibliography{Reference.bib} 

\begin{thebibliography}{109}
\expandafter\ifx\csname natexlab\endcsname\relax\def\natexlab#1{#1}\fi

\bibitem[{{Agra-Amboage} {et~al.}(2009){Agra-Amboage}, {Dougados}, {Cabrit},
  {Garcia}, \& {Ferruit}}]{Agra-Amboage2009}
{Agra-Amboage}, V., {Dougados}, C., {Cabrit}, S., {Garcia}, P.~J.~V., \&
  {Ferruit}, P. 2009, \aap, 493, 1029

\bibitem[{{Amara} {et~al.}(2015){Amara}, {Quanz}, \& {Akeret}}]{Amara2015}
{Amara}, A., {Quanz}, S.~P., \& {Akeret}, J. 2015, Astronomy and Computing, 10,
  107

\bibitem[{{Anglada} {et~al.}(2007){Anglada}, {L{\'o}pez}, {Estalella},
  {Masegosa}, {Riera}, \& {Raga}}]{Anglada2007}
{Anglada}, G., {L{\'o}pez}, R., {Estalella}, R., {et~al.} 2007, \aj, 133, 2799

\bibitem[{{Antoniucci} {et~al.}(2016){Antoniucci}, {Podio}, {Nisini},
  {Bacciotti}, {Lagadec}, {Sissa}, {La Camera}, {Giannini}, {Schmid},
  {Gratton}, {Turatto}, {Desidera}, {Bonnefoy}, {Chauvin}, {Dougados},
  {Bazzon}, {Thalmann}, \& {Langlois}}]{Antoniucci2016}
{Antoniucci}, S., {Podio}, L., {Nisini}, B., {et~al.} 2016, \aap, 593, L13

\bibitem[{{Arce} \& {Sargent}(2006)}]{Arce2006}
{Arce}, H.~G. \& {Sargent}, A.~I. 2006, \apj, 646, 1070

\bibitem[{{Armitage}(2015)}]{Armitage2015}
{Armitage}, P.~J. 2015, ArXiv e-prints

\bibitem[{{Avenhaus} {et~al.}(2014){Avenhaus}, {Quanz}, {Meyer}, {Brittain},
  {Carr}, \& {Najita}}]{Avenhaus2014b}
{Avenhaus}, H., {Quanz}, S.~P., {Meyer}, M.~R., {et~al.} 2014, \apj, 790, 56

\bibitem[{{Bacciotti} {et~al.}(2000){Bacciotti}, {Mundt}, {Ray},
  {Eisl{\"o}ffel}, {Solf}, \& {Camezind}}]{Bacciotti2000}
{Bacciotti}, F., {Mundt}, R., {Ray}, T.~P., {et~al.} 2000, \apj, 537, L49

\bibitem[{{Bacciotti} {et~al.}(2002){Bacciotti}, {Ray}, {Mundt},
  {Eisl{\"o}ffel}, \& {Solf}}]{Bacciotti2002}
{Bacciotti}, F., {Ray}, T.~P., {Mundt}, R., {Eisl{\"o}ffel}, J., \& {Solf}, J.
  2002, \apj, 576, 222

\bibitem[{{Bans} \& {K{\"o}nigl}(2012)}]{Bans2012}
{Bans}, A. \& {K{\"o}nigl}, A. 2012, \apj, 758, 100

\bibitem[{{Bertout} {et~al.}(1999){Bertout}, {Robichon}, \&
  {Arenou}}]{Bertout1999}
{Bertout}, C., {Robichon}, N., \& {Arenou}, F. 1999, \aap, 352, 574

\bibitem[{{Beuzit} {et~al.}(2019){Beuzit}, {Vigan}, {Mouillet}, {Dohlen},
  {Gratton}, {Boccaletti}, {Sauvage}, {Schmid}, {Langlois}, {Petit},
  {Baruffolo}, {Feldt}, {Milli}, {Wahhaj}, {Abe}, {Anselmi}, {Antichi},
  {Barette}, {Baudrand}, {Baudoz}, {Bazzon}, {Bernardi}, {Blanchard}, {Brast},
  {Bruno}, {Buey}, {Carbillet}, {Carle}, {Cascone}, {Chapron}, {Chauvin},
  {Charton}, {Claudi}, {Costille}, {De Caprio}, {Delboulb{\'e}}, {Desidera},
  {Dominik}, {Downing}, {Dupuis}, {Fabron}, {Fantinel}, {Farisato},
  {Feautrier}, {Fedrigo}, {Fusco}, {Gigan}, {Ginski}, {Girard}, {Giro},
  {Gisler}, {Gluck}, {Gry}, {Henning}, {Hubin}, {Hugot}, {Incorvaia}, {Jaquet},
  {Kasper}, {Lagadec}, {Lagrange}, {Coroller}, {Mignant}, {Ruyet}, {Lessio},
  {Lizon}, {Llored}, {Lundin}, {Madec}, {Magnard}, {Marteaud}, {Martinez},
  {Maurel}, {M{\'e}nard}, {Mesa}, {M{\"o}ller-Nilsson}, {Moulin}, {Moutou},
  {Orign{\'e}}, {Parisot}, {Pavlov}, {Perret}, {Pragt}, {Puget}, {Rabou},
  {Ramos}, {Reess}, {Rigal}, {Rochat}, {Roelfsema}, {Rousset}, {Roux},
  {Saisse}, {Salasnich}, {Santambrogio}, {Scuderi}, {Segransan}, {Sevin},
  {Siebenmorgen}, {Soenke}, {Stadler}, {Suarez}, {Tiph{\`e}ne}, {Turatto},
  {Udry}, {Vakili}, {Waters}, {Weber}, {Wildi}, {Zins}, \&
  {Zurlo}}]{Beuzit2019}
{Beuzit}, J.~L., {Vigan}, A., {Mouillet}, D., {et~al.} 2019, arXiv e-prints,
  arXiv:1902.04080

\bibitem[{{Blandford} \& {Payne}(1982)}]{Blandford1982}
{Blandford}, R.~D. \& {Payne}, D.~G. 1982, \mnras, 199, 883

\bibitem[{{Bonito} {et~al.}(2011){Bonito}, {Orlando}, {Miceli}, {Peres},
  {Micela}, \& {Favata}}]{Bonito2011}
{Bonito}, R., {Orlando}, S., {Miceli}, M., {et~al.} 2011, \apj, 737, 54

\bibitem[{{Bouvier} {et~al.}(1999){Bouvier}, {Chelli}, {Allain}, {Carrasco},
  {Costero}, {Cruz-Gonzalez}, {Dougados}, {Fern{\'a}ndez}, {Mart{\'{\i}}n},
  {M{\'e}nard}, {Mennessier}, {Mujica}, {Recillas}, {Salas}, {Schmidt}, \&
  {Wichmann}}]{Bouvier1999}
{Bouvier}, J., {Chelli}, A., {Allain}, S., {et~al.} 1999, \aap, 349, 619

\bibitem[{{Bressan} {et~al.}(2012){Bressan}, {Marigo}, {Girardi}, {Salasnich},
  {Dal Cero}, {Rubele}, \& {Nanni}}]{Bressan2012}
{Bressan}, A., {Marigo}, P., {Girardi}, L., {et~al.} 2012, \mnras, 427, 127

\bibitem[{{Cabrit}(2007)}]{Cabrit2007}
{Cabrit}, S. 2007, in IAU Symposium, Vol. 243, Star-Disk Interaction in Young
  Stars, ed. J.~{Bouvier} \& I.~{Appenzeller}, 203--214

\bibitem[{{Cabrit} {et~al.}(1990){Cabrit}, {Edwards}, {Strom}, \&
  {Strom}}]{Cabrit1990}
{Cabrit}, S., {Edwards}, S., {Strom}, S.~E., \& {Strom}, K.~M. 1990, \apj, 354,
  687

\bibitem[{{Calvet} {et~al.}(2004){Calvet}, {Muzerolle}, {Brice{\~n}o},
  {Hern{\'a}ndez}, {Hartmann}, {Saucedo}, \& {Gordon}}]{Calvet2004}
{Calvet}, N., {Muzerolle}, J., {Brice{\~n}o}, C., {et~al.} 2004, \aj, 128, 1294

\bibitem[{{Carbillet} {et~al.}(2011){Carbillet}, {Bendjoya}, {Abe}, {Guerri},
  {Boccaletti}, {Daban}, {Dohlen}, {Ferrari}, {Robbe-Dubois}, {Douet}, \&
  {Vakili}}]{Carbillet2011}
{Carbillet}, M., {Bendjoya}, P., {Abe}, L., {et~al.} 2011, Experimental
  Astronomy, 30, 39

\bibitem[{{Cardelli} {et~al.}(1989){Cardelli}, {Clayton}, \&
  {Mathis}}]{Cardelli1989}
{Cardelli}, J.~A., {Clayton}, G.~C., \& {Mathis}, J.~S. 1989, \apj, 345, 245

\bibitem[{{Chauvin} {et~al.}(2017){Chauvin}, {Desidera}, {Lagrange}, {Vigan},
  {Feldt}, {Gratton}, {Langlois}, {Cheetham}, {Bonnefoy}, \&
  {Meyer}}]{Chauvin2017}
{Chauvin}, G., {Desidera}, S., {Lagrange}, A.-M., {et~al.} 2017, in SF2A-2017:
  Proceedings of the Annual meeting of the French Society of Astronomy and
  Astrophysics, ed. C.~{Reyl{\'e}}, P.~{Di Matteo}, F.~{Herpin}, E.~{Lagadec},
  A.~{Lan{\c c}on}, Z.~{Meliani}, \& F.~{Royer}, 331--335

\bibitem[{{Claudi} {et~al.}(2008){Claudi}, {Turatto}, {Gratton}, {Antichi},
  {Bonavita}, {Bruno}, {Cascone}, {De Caprio}, {Desidera}, {Giro}, {Mesa},
  {Scuderi}, {Dohlen}, {Beuzit}, \& {Puget}}]{Claudi2008}
{Claudi}, R.~U., {Turatto}, M., {Gratton}, R.~G., {et~al.} 2008, in Society of
  Photo-Optical Instrumentation Engineers (SPIE) Conference Series, Vol. 7014,
  Society of Photo-Optical Instrumentation Engineers (SPIE) Conference Series,
  3

\bibitem[{{Coffey} {et~al.}(2004){Coffey}, {Bacciotti}, {Woitas}, {Ray}, \&
  {Eisl{\"o}ffel}}]{Coffey2004}
{Coffey}, D., {Bacciotti}, F., {Woitas}, J., {Ray}, T.~P., \& {Eisl{\"o}ffel},
  J. 2004, \apss, 292, 553

\bibitem[{{Coffey} {et~al.}(2015){Coffey}, {Dougados}, {Cabrit}, {Pety}, \&
  {Bacciotti}}]{Coffey2015}
{Coffey}, D., {Dougados}, C., {Cabrit}, S., {Pety}, J., \& {Bacciotti}, F.
  2015, \apj, 804, 2

\bibitem[{{Corcoran} \& {Ray}(1997)}]{Corcoran1997}
{Corcoran}, M. \& {Ray}, T.~P. 1997, \aap, 321, 189

\bibitem[{{Delorme} {et~al.}(2017){Delorme}, {Meunier}, {Albert}, {Lagadec},
  {Le Coroller}, {Galicher}, {Mouillet}, {Boccaletti}, {Mesa}, {Meunier},
  {Beuzit}, {Lagrange}, {Chauvin}, {Sapone}, {Langlois}, {Maire},
  {Montarg{\`e}s}, {Gratton}, {Vigan}, \& {Surace}}]{Delorme2017}
{Delorme}, P., {Meunier}, N., {Albert}, D., {et~al.} 2017, in SF2A-2017:
  Proceedings of the Annual meeting of the French Society of Astronomy and
  Astrophysics, ed. C.~{Reyl{\'e}}, P.~{Di Matteo}, F.~{Herpin}, E.~{Lagadec},
  A.~{Lan{\c c}on}, Z.~{Meliani}, \& F.~{Royer}, 347--361

\bibitem[{{Devine} {et~al.}(2000){Devine}, {Grady}, {Kimble}, {Woodgate},
  {Bruhweiler}, {Boggess}, {Linsky}, \& {Clampin}}]{Devine2000}
{Devine}, D., {Grady}, C.~A., {Kimble}, R.~A., {et~al.} 2000, \apjl, 542, L115

\bibitem[{{Dohlen} {et~al.}(2008){Dohlen}, {Langlois}, {Saisse}, {Hill},
  {Origne}, {Jacquet}, {Fabron}, {Blanc}, {Llored}, {Carle}, {Moutou}, {Vigan},
  {Boccaletti}, {Carbillet}, {Mouillet}, \& {Beuzit}}]{Dohlen2008}
{Dohlen}, K., {Langlois}, M., {Saisse}, M., {et~al.} 2008, in Society of
  Photo-Optical Instrumentation Engineers (SPIE) Conference Series, Vol. 7014,
  Society of Photo-Optical Instrumentation Engineers (SPIE) Conference Series,
  3

\bibitem[{{Dotter}(2016)}]{Dotter2016}
{Dotter}, A. 2016, \apjs, 222, 8

\bibitem[{{Dotter} {et~al.}(2008){Dotter}, {Chaboyer}, {Jevremovi{\'c}},
  {Kostov}, {Baron}, \& {Ferguson}}]{Dotter2008}
{Dotter}, A., {Chaboyer}, B., {Jevremovi{\'c}}, D., {et~al.} 2008, \apjs, 178,
  89

\bibitem[{{Edwards} {et~al.}(2006){Edwards}, {Fischer}, {Hillenbrand}, \&
  {Kwan}}]{Edwards2006}
{Edwards}, S., {Fischer}, W., {Hillenbrand}, L., \& {Kwan}, J. 2006, \apj, 646,
  319

\bibitem[{{Eiroa} {et~al.}(2000){Eiroa}, {Mora}, {Palacios}, {Alberdi},
  {Miranda}, {Cameron}, {Horne}, {Tsapras}, {Davies}, {Deeg}, {Garz{\'o}n}, {de
  Winter}, {Ferlet}, {Grady}, {Harris}, {Rauer}, {Mer{\'{\i}}n}, {Montesinos},
  {Solano}, {Oudmaijer}, {Penny}, {Quirrenbach}, {Schneider}, \&
  {Wesselius}}]{Eiroa2000}
{Eiroa}, C., {Mora}, A., {Palacios}, J., {et~al.} 2000, in Astronomical Society
  of the Pacific Conference Series, Vol. 219, Disks, Planetesimals, and
  Planets, ed. G.~{Garz{\'o}n}, C.~{Eiroa}, D.~{de Winter}, \& T.~J. {Mahoney},
  3

\bibitem[{{Eisloffel} {et~al.}(2000){Eisloffel}, {Mundt}, {Ray}, \&
  {Rodriguez}}]{Eisloffel2000}
{Eisloffel}, J., {Mundt}, R., {Ray}, T.~P., \& {Rodriguez}, L.~F. 2000, in
  Protostars and Planets IV, ed. V.~{Mannings}, A.~P. {Boss}, \& S.~S.
  {Russell}, 815

\bibitem[{{Ellerbroek} {et~al.}(2014){Ellerbroek}, {Podio}, {Dougados},
  {Cabrit}, {Sitko}, {Sana}, {Kaper}, {de Koter}, {Klaassen}, {Mulders},
  {Mendigut{\'{\i}}a}, {Grady}, {Grankin}, {van Winckel}, {Bacciotti},
  {Russell}, {Lynch}, {Hammel}, {Beerman}, {Day}, {Huelsman}, {Werren},
  {Henden}, \& {Grindlay}}]{Ellerbroek2014}
{Ellerbroek}, L.~E., {Podio}, L., {Dougados}, C., {et~al.} 2014, \aap, 563, A87

\bibitem[{{ESA}(1997)}]{Esa1997}
{ESA}, ed. 1997, ESA Special Publication, Vol. 1200, {The HIPPARCOS and TYCHO
  catalogues. Astrometric and photometric star catalogues derived from the ESA
  HIPPARCOS Space Astrometry Mission}

\bibitem[{{Facchini} {et~al.}(2018){Facchini}, {Juh{\'a}sz}, \&
  {Lodato}}]{Facchini2018}
{Facchini}, S., {Juh{\'a}sz}, A., \& {Lodato}, G. 2018, \mnras, 473, 4459

\bibitem[{{Frank} {et~al.}(2014){Frank}, {Ray}, {Cabrit}, {Hartigan}, {Arce},
  {Bacciotti}, {Bally}, {Benisty}, {Eisl{\"o}ffel}, {G{\"u}del}, {Lebedev},
  {Nisini}, \& {Raga}}]{Frank2014}
{Frank}, A., {Ray}, T.~P., {Cabrit}, S., {et~al.} 2014, Protostars and Planets
  VI, 451

\bibitem[{{Fusco} {et~al.}(2006){Fusco}, {Petit}, {Rousset}, {Sauvage},
  {Dohlen}, {Mouillet}, {Charton}, {Baudoz}, {Kasper}, {Fedrigo}, {Rabou},
  {Feautrier}, {Downing}, {Gigan}, {Conan}, {Beuzit}, {Hubin}, {Wildi}, \&
  {Puget}}]{Fusco2006}
{Fusco}, T., {Petit}, C., {Rousset}, G., {et~al.} 2006, in Society of
  Photo-Optical Instrumentation Engineers (SPIE) Conference Series, Vol. 6272,
  Society of Photo-Optical Instrumentation Engineers (SPIE) Conference Series,
  0

\bibitem[{{Gaia Collaboration}(2016)}]{GAIA2016}
{Gaia Collaboration}. 2016, VizieR Online Data Catalog, 1337

\bibitem[{{Gaia Collaboration} {et~al.}(2018){Gaia Collaboration}, {Brown},
  {Vallenari}, {Prusti}, {de Bruijne}, {Babusiaux}, \&
  {Bailer-Jones}}]{GAIA2018}
{Gaia Collaboration}, {Brown}, A.~G.~A., {Vallenari}, A., {et~al.} 2018, ArXiv
  e-prints [\eprint[arXiv]{1804.09365}]

\bibitem[{{Galicher} {et~al.}(2018){Galicher}, {Boccaletti}, {Mesa}, {Delorme},
  {Gratton}, {Langlois}, {Lagrange}, {Maire}, {Le Coroller}, {Chauvin},
  {Biller}, {Cantalloube}, {Janson}, {Lagadec}, {Meunier}, {Vigan},
  {Hagelberg}, {Bonnefoy}, {Zurlo}, {Rocha}, {Maurel}, {Jaquet}, {Buey}, \&
  {Weber}}]{Galicher2018}
{Galicher}, R., {Boccaletti}, A., {Mesa}, D., {et~al.} 2018, \aap, 615, A92

\bibitem[{{Garufi} {et~al.}(2018){Garufi}, {Benisty}, {Pinilla}, {Tazzari},
  {Dominik}, {Ginski}, {Henning}, {Kral}, {Langlois}, {M{\'e}nard}, {Stolker},
  {Szulagyi}, {Villenave}, \& {van der Plas}}]{Garufi2018}
{Garufi}, A., {Benisty}, M., {Pinilla}, P., {et~al.} 2018, \aap, 620, A94

\bibitem[{{Garufi} {et~al.}(2017){Garufi}, {Meeus}, {Benisty}, {Quanz},
  {Banzatti}, {Kama}, {Canovas}, {Eiroa}, {Schmid}, {Stolker}, {Pohl},
  {Rigliaco}, {M{\'e}nard}, {Meyer}, {van Boekel}, \& {Dominik}}]{Garufi2017}
{Garufi}, A., {Meeus}, G., {Benisty}, M., {et~al.} 2017, \aap, 603, A21

\bibitem[{{Garufi} {et~al.}(2016){Garufi}, {Quanz}, {Schmid}, {Mulders},
  {Avenhaus}, {Boccaletti}, {Ginski}, {Langlois}, {Stolker}, {Augereau},
  {Benisty}, {Lopez}, {Dominik}, {Gratton}, {Henning}, {Janson}, {M{\'e}nard},
  {Meyer}, {Pinte}, {Sissa}, {Vigan}, {Zurlo}, {Bazzon}, {Buenzli}, {Bonnefoy},
  {Brandner}, {Chauvin}, {Cheetham}, {Cudel}, {Desidera}, {Feldt}, {Galicher},
  {Kasper}, {Lagrange}, {Lannier}, {Maire}, {Mesa}, {Mouillet}, {Peretti},
  {Perrot}, {Salter}, \& {Wildi}}]{Garufi2016}
{Garufi}, A., {Quanz}, S.~P., {Schmid}, H.~M., {et~al.} 2016, \aap, 588, A8

\bibitem[{{G{\'o}mez de Castro} \& {Verdugo}(2001)}]{GomezdeCastro2001}
{G{\'o}mez de Castro}, A.~I. \& {Verdugo}, E. 2001, \apj, 548, 976

\bibitem[{{Grady} {et~al.}(2010){Grady}, {Hamaguchi}, {Schneider}, {Stecklum},
  {Woodgate}, {McCleary}, {Williger}, {Sitko}, {M{\'e}nard}, {Henning},
  {Brittain}, {Troutmann}, {Donehew}, {Hines}, {Wisniewski}, {Lynch},
  {Russell}, {Rudy}, {Day}, {Shenoy}, {Wilner}, {Silverstone}, {Bouret},
  {Meusinger}, {Clampin}, {Kim}, {Petre}, {Sahu}, {Endres}, \&
  {Collins}}]{Grady2010}
{Grady}, C.~A., {Hamaguchi}, K., {Schneider}, G., {et~al.} 2010, \apj, 719,
  1565

\bibitem[{{Grady} {et~al.}(2004){Grady}, {Woodgate}, {Torres}, {Henning},
  {Apai}, {Rodmann}, {Wang}, {Stecklum}, {Linz}, {Williger}, {Brown},
  {Wilkinson}, {Harper}, {Herczeg}, {Danks}, {Vieira}, {Malumuth}, {Collins},
  \& {Hill}}]{Grady2004}
{Grady}, C.~A., {Woodgate}, B., {Torres}, C.~A.~O., {et~al.} 2004, \apj, 608,
  809

\bibitem[{{Gregory} {et~al.}(2012){Gregory}, {Donati}, {Morin}, {Hussain},
  {Mayne}, {Hillenbrand}, \& {Jardine}}]{Gregory2012}
{Gregory}, S.~G., {Donati}, J.-F., {Morin}, J., {et~al.} 2012, \apj, 755, 97

\bibitem[{{Hartigan} {et~al.}(1995){Hartigan}, {Edwards}, \&
  {Ghandour}}]{Hartigan1995}
{Hartigan}, P., {Edwards}, S., \& {Ghandour}, L. 1995, \apj, 452, 736

\bibitem[{{Hartigan} \& {Morse}(2007)}]{Hartigan2007}
{Hartigan}, P. \& {Morse}, J. 2007, \apj, 660, 426

\bibitem[{{Hartigan} {et~al.}(1994){Hartigan}, {Morse}, \&
  {Raymond}}]{Hartigan1994}
{Hartigan}, P., {Morse}, J.~A., \& {Raymond}, J. 1994, \apj, 436, 125

\bibitem[{{Hartigan} {et~al.}(1987){Hartigan}, {Raymond}, \&
  {Hartmann}}]{Hartigan1987}
{Hartigan}, P., {Raymond}, J., \& {Hartmann}, L. 1987, \apj, 316, 323

\bibitem[{{Herbst} {et~al.}(1994){Herbst}, {Herbst}, {Grossman}, \&
  {Weinstein}}]{Herbst1994}
{Herbst}, W., {Herbst}, D.~K., {Grossman}, E.~J., \& {Weinstein}, D. 1994, \aj,
  108, 1906

\bibitem[{{Herbst} \& {Shevchenko}(1999)}]{Herbst1999}
{Herbst}, W. \& {Shevchenko}, V.~S. 1999, \aj, 118, 1043

\bibitem[{{Hubrig} {et~al.}(2018){Hubrig}, {Jarvinen}, {Sch{\"o}ller},
  {Carroll}, {Ilyin}, \& {Pogodin}}]{Hubrig2018}
{Hubrig}, S., {Jarvinen}, S.~P., {Sch{\"o}ller}, M., {et~al.} 2018, arXiv
  e-prints [\eprint[arXiv]{1812.04482}]

\bibitem[{{Isella} {et~al.}(2010){Isella}, {Carpenter}, \&
  {Sargent}}]{Isella2010}
{Isella}, A., {Carpenter}, J.~M., \& {Sargent}, A.~I. 2010, \apj, 714, 1746

\bibitem[{{Kenyon} {et~al.}(1994){Kenyon}, {Gomez}, {Marzke}, \&
  {Hartmann}}]{Kenyon1994}
{Kenyon}, S.~J., {Gomez}, M., {Marzke}, R.~O., \& {Hartmann}, L. 1994, \aj,
  108, 251

\bibitem[{{Kenyon} \& {Hartmann}(1995)}]{Kenyon1995}
{Kenyon}, S.~J. \& {Hartmann}, L. 1995, \apjs, 101, 117

\bibitem[{{Konigl} \& {Pudritz}(2000)}]{Konigl2000}
{Konigl}, A. \& {Pudritz}, R.~E. 2000, Protostars and Planets IV, 759

\bibitem[{{Kurucz}(1993)}]{Kurucz1993}
{Kurucz}, R. 1993, SYNTHE Spectrum Synthesis Programs and Line Data.~Kurucz
  CD-ROM No.~18.~Cambridge, Mass.: Smithsonian Astrophysical Observatory,
  1993., 18

\bibitem[{{Lada}(1987)}]{Lada1987}
{Lada}, C.~J. 1987, in IAU Symposium, Vol. 115, Star Forming Regions, ed.
  M.~{Peimbert} \& J.~{Jugaku}, 1--17

\bibitem[{{Lindegren} {et~al.}(2018){Lindegren}, {Hern{\'a}ndez}, {Bombrun},
  {Klioner}, {Bastian}, {Ramos-Lerate}, {de Torres}, {Steidelm{\"u}ller},
  {Stephenson}, {Hobbs}, {Lammers}, {Biermann}, {Geyer}, {Hilger}, {Michalik},
  {Stampa}, {McMillan}, {Casta{\~n}eda}, {Clotet}, {Comoretto}, {Davidson},
  {Fabricius}, {Gracia}, {Hambly}, {Hutton}, {Mora}, {Portell}, {van Leeuwen},
  {Abbas}, {Abreu}, {Altmann}, {Andrei}, {Anglada}, {Balaguer-N{\'u}{\~n}ez},
  {Barache}, {Becciani}, {Bertone}, {Bianchi}, {Bouquillon}, {Bourda},
  {Br{\"u}semeister}, {Bucciarelli}, {Busonero}, {Buzzi}, {Cancelliere},
  {Carlucci}, {Charlot}, {Cheek}, {Crosta}, {Crowley}, {de Bruijne}, {de
  Felice}, {Drimmel}, {Esquej}, {Fienga}, {Fraile}, {Gai}, {Garralda},
  {Gonz{\'a}lez-Vidal}, {Guerra}, {Hauser}, {Hofmann}, {Holl}, {Jordan},
  {Lattanzi}, {Lenhardt}, {Liao}, {Licata}, {Lister}, {L{\"o}ffler},
  {Marchant}, {Martin-Fleitas}, {Messineo}, {Mignard}, {Morbidelli}, {Poggio},
  {Riva}, {Rowell}, {Salguero}, {Sarasso}, {Sciacca}, {Siddiqui}, {Smart},
  {Spagna}, {Steele}, {Taris}, {Torra}, {van Elteren}, {van Reeven}, \&
  {Vecchiato}}]{Lindegren2018}
{Lindegren}, L., {Hern{\'a}ndez}, J., {Bombrun}, A., {et~al.} 2018, \aap, 616,
  A2

\bibitem[{{Long} {et~al.}(2018){Long}, {Pinilla}, {Herczeg}, {Harsono},
  {Dipierro}, {Pascucci}, {Hendler}, {Tazzari}, {Ragusa}, {Salyk}, {Edwards},
  {Lodato}, {van de Plas}, {Johnstone}, {Liu}, {Boehler}, {Cabrit}, {Manara},
  {Menard}, {Mulders}, {Nisini}, {Fischer}, {Rigliaco}, {Banzatti}, {Avenhaus},
  \& {Gully-Santiago}}]{Long2018b}
{Long}, F., {Pinilla}, P., {Herczeg}, G.~J., {et~al.} 2018, \apj, 869, 17

\bibitem[{{Luhman}(2018)}]{Luhman2018}
{Luhman}, K.~L. 2018, ArXiv e-prints [\eprint[arXiv]{1811.01359}]

\bibitem[{{Maire} {et~al.}(2016){Maire}, {Bonnefoy}, {Ginski}, {Vigan},
  {Messina}, {Mesa}, {Galicher}, {Gratton}, {Desidera}, {Kopytova}, {Millward},
  {Thalmann}, {Claudi}, {Ehrenreich}, {Zurlo}, {Chauvin}, {Antichi},
  {Baruffolo}, {Bazzon}, {Beuzit}, {Blanchard}, {Boccaletti}, {de Boer},
  {Carle}, {Cascone}, {Costille}, {De Caprio}, {Delboulb{\'e}}, {Dohlen},
  {Dominik}, {Feldt}, {Fusco}, {Girard}, {Giro}, {Gisler}, {Gluck}, {Gry},
  {Henning}, {Hubin}, {Hugot}, {Jaquet}, {Kasper}, {Lagrange}, {Langlois}, {Le
  Mignant}, {Llored}, {Madec}, {Martinez}, {Mawet}, {Milli},
  {M{\"o}ller-Nilsson}, {Mouillet}, {Moulin}, {Moutou}, {Orign{\'e}}, {Pavlov},
  {Petit}, {Pragt}, {Puget}, {Ramos}, {Rochat}, {Roelfsema}, {Salasnich},
  {Sauvage}, {Schmid}, {Turatto}, {Udry}, {Vakili}, {Wahhaj}, {Weber}, \&
  {Wildi}}]{Maire2016}
{Maire}, A.~L., {Bonnefoy}, M., {Ginski}, C., {et~al.} 2016, \aap, 587, A56

\bibitem[{{Marois} {et~al.}(2014){Marois}, {Correia}, {V{\'e}ran}, \&
  {Currie}}]{Marois2014}
{Marois}, C., {Correia}, C., {V{\'e}ran}, J.-P., \& {Currie}, T. 2014, in IAU
  Symposium, Vol. 299, Exploring the Formation and Evolution of Planetary
  Systems, ed. M.~{Booth}, B.~C. {Matthews}, \& J.~R. {Graham}, 48--49

\bibitem[{{Marois} {et~al.}(2006){Marois}, {Lafreni{\`e}re}, {Doyon},
  {Macintosh}, \& {Nadeau}}]{Marois2006}
{Marois}, C., {Lafreni{\`e}re}, D., {Doyon}, R., {Macintosh}, B., \& {Nadeau},
  D. 2006, \apj, 641, 556

\bibitem[{{Mendigut{\'{\i}}a} {et~al.}(2011){Mendigut{\'{\i}}a}, {Calvet},
  {Montesinos}, {Mora}, {Muzerolle}, {Eiroa}, {Oudmaijer}, \&
  {Mer{\'{\i}}n}}]{Mendigutia2011}
{Mendigut{\'{\i}}a}, I., {Calvet}, N., {Montesinos}, B., {et~al.} 2011, \aap,
  535, A99

\bibitem[{{Mesa} {et~al.}(2019){Mesa}, {Bonnefoy}, {Gratton}, {Van Der Plas},
  {D'Orazi}, {Sissa}, {Zurlo}, {Rigliaco}, {Schmidt}, {Langlois}, {Vigan},
  {Ubeira Gabellini}, {Desidera}, {Antoniucci}, {Barbieri}, {Benisty},
  {Boccaletti}, {Claudi}, {Fedele}, {Gasparri}, {Henning}, {Kasper},
  {Lagrange}, {Lazzoni}, {Lodato}, {Maire}, {Manara}, {Meyer}, {Reggiani},
  {Samland}, {Van den Ancker}, {Chauvin}, {Cheetham}, {Feldt}, {Hugot},
  {Janson}, {Ligi}, {Moller-Nilsson}, {Petit}, {Rickman}, {Rigal}, \&
  {Wildi}}]{Mesa2019}
{Mesa}, D., {Bonnefoy}, M., {Gratton}, R., {et~al.} 2019, arXiv e-prints
  [\eprint[arXiv]{1902.02536}]

\bibitem[{{Min} {et~al.}(2017){Min}, {Stolker}, {Dominik}, \&
  {Benisty}}]{Min2017}
{Min}, M., {Stolker}, T., {Dominik}, C., \& {Benisty}, M. 2017, \aap, 604, L10

\bibitem[{{Mora} {et~al.}(2001){Mora}, {Mer{\'{\i}}n}, {Solano}, {Montesinos},
  {de Winter}, {Eiroa}, {Ferlet}, {Grady}, {Davies}, {Miranda}, {Oudmaijer},
  {Palacios}, {Quirrenbach}, {Harris}, {Rauer}, {Collier Cameron}, {Deeg},
  {Garz{\'o}n}, {Penny}, {Schneider}, {Tsapras}, \& {Wesselius}}]{Mora2001}
{Mora}, A., {Mer{\'{\i}}n}, B., {Solano}, E., {et~al.} 2001, \aap, 378, 116

\bibitem[{{Nakajima} \& {Golimowski}(1995)}]{Nakajima1995}
{Nakajima}, T. \& {Golimowski}, D.~A. 1995, \aj, 109, 1181

\bibitem[{{Nguyen} {et~al.}(2012){Nguyen}, {Brandeker}, {van Kerkwijk}, \&
  {Jayawardhana}}]{Nguyen2012}
{Nguyen}, D.~C., {Brandeker}, A., {van Kerkwijk}, M.~H., \& {Jayawardhana}, R.
  2012, \apj, 745, 119

\bibitem[{{Nisini} {et~al.}(2005){Nisini}, {Bacciotti}, {Giannini}, {Massi},
  {Eisl{\"o}ffel}, {Podio}, \& {Ray}}]{Nisini2005}
{Nisini}, B., {Bacciotti}, F., {Giannini}, T., {et~al.} 2005, \aap, 441, 159

\bibitem[{{Nisini} {et~al.}(2015){Nisini}, {Santangelo}, {Giannini},
  {Antoniucci}, {Cabrit}, {Codella}, {Davis}, {Eisl{\"o}ffel}, {Kristensen},
  {Herczeg}, {Neufeld}, \& {van Dishoeck}}]{Nisini2015}
{Nisini}, B., {Santangelo}, G., {Giannini}, T., {et~al.} 2015, \apj, 801, 121

\bibitem[{{Oudmaijer} {et~al.}(2001){Oudmaijer}, {Palacios}, {Eiroa}, {Davies},
  {de Winter}, {Ferlet}, {Garz{\'o}n}, {Grady}, {Collier Cameron}, {Deeg},
  {Harris}, {Horne}, {Mer{\'{\i}}n}, {Miranda}, {Montesinos}, {Mora}, {Penny},
  {Quirrenbach}, {Rauer}, {Schneider}, {Solano}, {Tsapras}, \&
  {Wesselius}}]{Oudmaijer2001}
{Oudmaijer}, R.~D., {Palacios}, J., {Eiroa}, C., {et~al.} 2001, \aap, 379, 564

\bibitem[{{Pavlov} {et~al.}(2008){Pavlov}, {Feldt}, \& {Henning}}]{Pavlov2008}
{Pavlov}, A., {Feldt}, M., \& {Henning}, T. 2008, in Astronomical Society of
  the Pacific Conference Series, Vol. 394, Astronomical Data Analysis Software
  and Systems XVII, ed. R.~W. {Argyle}, P.~S. {Bunclark}, \& J.~R. {Lewis}, 581

\bibitem[{{Petrov} {et~al.}(1999){Petrov}, {Zajtseva}, {Efimov}, {Duemmler},
  {Ilyin}, {Tuominen}, \& {Shcherbakov}}]{Petrov1999}
{Petrov}, P.~P., {Zajtseva}, G.~V., {Efimov}, Y.~S., {et~al.} 1999, \aap, 341,
  553

\bibitem[{{Pinilla} {et~al.}(2012){Pinilla}, {Benisty}, \&
  {Birnstiel}}]{Pinilla2012}
{Pinilla}, P., {Benisty}, M., \& {Birnstiel}, T. 2012, \aap, 545, A81

\bibitem[{{Pinilla} {et~al.}(2018){Pinilla}, {Tazzari}, {Pascucci}, {Youdin},
  {Garufi}, {Manara}, {Testi}, {van der Plas}, {Barenfeld}, {Canovas}, {Cox},
  {Hendler}, {P{\'e}rez}, \& {van der Marel}}]{Pinilla2018}
{Pinilla}, P., {Tazzari}, M., {Pascucci}, I., {et~al.} 2018, \apj, 859, 32

\bibitem[{{Podio} {et~al.}(2011){Podio}, {Eisl{\"o}ffel}, {Melnikov}, {Hodapp},
  \& {Bacciotti}}]{Podio2011}
{Podio}, L., {Eisl{\"o}ffel}, J., {Melnikov}, S., {Hodapp}, K.~W., \&
  {Bacciotti}, F. 2011, \aap, 527, A13

\bibitem[{{Pohl} {et~al.}(2017){Pohl}, {Benisty}, {Pinilla}, {Ginski}, {de
  Boer}, {Avenhaus}, {Henning}, {Zurlo}, {Boccaletti}, {Augereau}, {Birnstiel},
  {Dominik}, {Facchini}, {Fedele}, {Janson}, {Keppler}, {Kral}, {Langlois},
  {Ligi}, {Maire}, {M{\'e}nard}, {Meyer}, {Pinte}, {Quanz}, {Sauvage},
  {Sezestre}, {Stolker}, {Szul{\'a}gyi}, {van Boekel}, {van der Plas},
  {Villenave}, {Baruffolo}, {Baudoz}, {Le Mignant}, {Maurel}, {Ramos}, \&
  {Weber}}]{Pohl2017}
{Pohl}, A., {Benisty}, M., {Pinilla}, P., {et~al.} 2017, \apj, 850, 52

\bibitem[{{Pudritz} {et~al.}(2007){Pudritz}, {Ouyed}, {Fendt}, \&
  {Brandenburg}}]{Pudritz2007}
{Pudritz}, R.~E., {Ouyed}, R., {Fendt}, C., \& {Brandenburg}, A. 2007,
  Protostars and Planets V, 277

\bibitem[{{Quanz} {et~al.}(2011){Quanz}, {Schmid}, {Geissler}, {Meyer},
  {Henning}, {Brandner}, \& {Wolf}}]{Quanz2011}
{Quanz}, S.~P., {Schmid}, H.~M., {Geissler}, K., {et~al.} 2011, \apj, 738, 23

\bibitem[{{Raga} {et~al.}(1998){Raga}, {Canto}, \& {Cabrit}}]{Raga1998}
{Raga}, A.~C., {Canto}, J., \& {Cabrit}, S. 1998, \aap, 332, 714

\bibitem[{{Ray} {et~al.}(2007){Ray}, {Dougados}, {Bacciotti}, {Eisl{\"o}ffel},
  \& {Chrysostomou}}]{Ray2007}
{Ray}, T., {Dougados}, C., {Bacciotti}, F., {Eisl{\"o}ffel}, J., \&
  {Chrysostomou}, A. 2007, Protostars and Planets V, 231

\bibitem[{{Robitaille} {et~al.}(2007){Robitaille}, {Whitney}, {Indebetouw}, \&
  {Wood}}]{Robitaille2007}
{Robitaille}, T.~P., {Whitney}, B.~A., {Indebetouw}, R., \& {Wood}, K. 2007,
  \apjs, 169, 328

\bibitem[{{Rubinstein} {et~al.}(2018){Rubinstein}, {Mac{\'{\i}}as},
  {Espaillat}, {Zhang}, {Calvet}, \& {Robinson}}]{Rubinstein2018}
{Rubinstein}, A.~E., {Mac{\'{\i}}as}, E., {Espaillat}, C.~C., {et~al.} 2018,
  \apj, 860, 7

\bibitem[{{Sauty} \& {Tsinganos}(1994)}]{Sauty1994}
{Sauty}, C. \& {Tsinganos}, K. 1994, \aap, 287, 893

\bibitem[{{Sbordone} {et~al.}(2004){Sbordone}, {Bonifacio}, {Castelli}, \&
  {Kurucz}}]{Sbordone2004}
{Sbordone}, L., {Bonifacio}, P., {Castelli}, F., \& {Kurucz}, R.~L. 2004,
  Memorie della Societa Astronomica Italiana Supplementi, 5, 93

\bibitem[{{Schegerer} {et~al.}(2008){Schegerer}, {Wolf}, {Ratzka}, \&
  {Leinert}}]{Schegerer2008}
{Schegerer}, A.~A., {Wolf}, S., {Ratzka}, T., \& {Leinert}, C. 2008, \aap, 478,
  779

\bibitem[{{Schmid} {et~al.}(2017){Schmid}, {Bazzon}, {Milli}, {Roelfsema},
  {Engler}, {Mouillet}, {Lagadec}, {Sissa}, {Sauvage}, {Ginski}, {Baruffolo},
  {Beuzit}, {Boccaletti}, {Bohn}, {Claudi}, {Costille}, {Desidera}, {Dohlen},
  {Dominik}, {Feldt}, {Fusco}, {Gisler}, {Girard}, {Gratton}, {Henning},
  {Hubin}, {Joos}, {Kasper}, {Langlois}, {Pavlov}, {Pragt}, {Puget}, {Quanz},
  {Salasnich}, {Siebenmorgen}, {Stute}, {Suarez}, {Szul{\'a}gyi}, {Thalmann},
  {Turatto}, {Udry}, {Vigan}, \& {Wildi}}]{Schmid2017}
{Schmid}, H.~M., {Bazzon}, A., {Milli}, J., {et~al.} 2017, \aap, 602, A53

\bibitem[{{Schmid} {et~al.}(2018){Schmid}, {Bazzon}, {Roelfsema}, {Mouillet},
  {Milli}, {Menard}, {Gisler}, {Hunziker}, {Pragt}, {Dominik}, {Boccaletti},
  {Ginski}, {Abe}, {Antoniucci}, {Avenhaus}, {Baruffolo}, {Baudoz}, {Beuzit},
  {Carbillet}, {Chauvin}, {Claudi}, {Costille}, {Daban}, {de Haan}, {Desidera},
  {Dohlen}, {Downing}, {Elswijk}, {Engler}, {Feldt}, {Fusco}, {Girard},
  {Gratton}, {Hanenburg}, {Henning}, {Hubin}, {Joos}, {Kasper}, {Keller},
  {Langlois}, {Lagadec}, {Martinez}, {Mulder}, {Pavlov}, {Podio}, {Puget},
  {Quanz}, {Rigal}, {Salasnich}, {Sauvage}, {Schuil}, {Siebenmorgen}, {Sissa},
  {Snik}, {Suarez}, {Thalmann}, {Turatto}, {Udry}, {van Duin}, {van Holstein},
  {Vigan}, \& {Wildi}}]{Schmid2018}
{Schmid}, H.~M., {Bazzon}, A., {Roelfsema}, R., {et~al.} 2018, \aap, 619, A9

\bibitem[{{Schmid} {et~al.}(2006){Schmid}, {Joos}, \& {Tschan}}]{Schmid2006}
{Schmid}, H.~M., {Joos}, F., \& {Tschan}, D. 2006, \aap, 452, 657

\bibitem[{{Seale} \& {Looney}(2008)}]{Seale2008}
{Seale}, J.~P. \& {Looney}, L.~W. 2008, \apj, 675, 427

\bibitem[{{Sheehan} \& {Eisner}(2017)}]{Sheehan2017}
{Sheehan}, P.~D. \& {Eisner}, J.~A. 2017, \apjl, 840, L12

\bibitem[{{Sitko} {et~al.}(2008){Sitko}, {Carpenter}, {Kimes}, {Wilde},
  {Lynch}, {Russell}, {Rudy}, {Mazuk}, {Venturini}, {Puetter}, {Grady},
  {Polomski}, {Wisnewski}, {Brafford}, {Hammel}, \& {Perry}}]{Sitko2008}
{Sitko}, M.~L., {Carpenter}, W.~J., {Kimes}, R.~L., {et~al.} 2008, \apj, 678,
  1070

\bibitem[{{Skinner} {et~al.}(2011){Skinner}, {Audard}, \&
  {G{\"u}del}}]{Skinner2011}
{Skinner}, S.~L., {Audard}, M., \& {G{\"u}del}, M. 2011, \apj, 737, 19

\bibitem[{{Skinner} {et~al.}(2018){Skinner}, {Schneider}, {Audard}, \&
  {G{\"u}del}}]{Skinner2018}
{Skinner}, S.~L., {Schneider}, P.~C., {Audard}, M., \& {G{\"u}del}, M. 2018,
  \apj, 855, 143

\bibitem[{{Soummer} {et~al.}(2012){Soummer}, {Pueyo}, \&
  {Larkin}}]{Soummer2012}
{Soummer}, R., {Pueyo}, L., \& {Larkin}, J. 2012, \apjl, 755, L28

\bibitem[{{St-Onge} \& {Bastien}(2008)}]{St-Onge2008}
{St-Onge}, G. \& {Bastien}, P. 2008, \apj, 674, 1032

\bibitem[{{Takami} {et~al.}(2007){Takami}, {Beck}, {Pyo}, {McGregor}, \&
  {Davis}}]{Takami2007}
{Takami}, M., {Beck}, T.~L., {Pyo}, T.-S., {McGregor}, P., \& {Davis}, C. 2007,
  \apj, 670, L33

\bibitem[{{Takami} {et~al.}(2002){Takami}, {Chrysostomou}, {Bailey},
  {Gledhill}, {Tamura}, \& {Terada}}]{Takami2002}
{Takami}, M., {Chrysostomou}, A., {Bailey}, J., {et~al.} 2002, \apjl, 568, L53

\bibitem[{{Takami} {et~al.}(2013){Takami}, {Karr}, {Hashimoto}, {Kim},
  {Wisniewski}, {Henning}, {Grady}, {Kandori}, {Hodapp}, {Kudo}, {Kusakabe},
  {Chou}, {Itoh}, {Momose}, {Mayama}, {Currie}, {Follette}, {Kwon}, {Abe},
  {Brandner}, {Brandt}, {Carson}, {Egner}, {Feldt}, {Guyon}, {Hayano},
  {Hayashi}, {Hayashi}, {Ishii}, {Iye}, {Janson}, {Knapp}, {Kuzuhara},
  {McElwain}, {Matsuo}, {Miyama}, {Morino}, {Moro-Martin}, {Nishimura}, {Pyo},
  {Serabyn}, {Suto}, {Suzuki}, {Takato}, {Terada}, {Thalmann}, {Tomono},
  {Turner}, {Watanabe}, {Yamada}, {Takami}, {Usuda}, \& {Tamura}}]{Takami2013}
{Takami}, M., {Karr}, J.~L., {Hashimoto}, J., {et~al.} 2013, \apj, 772, 145

\bibitem[{{Varga} {et~al.}(2018){Varga}, {{\'A}brah{\'a}m}, {Chen}, {Ratzka},
  {Gab{\'a}nyi}, {K{\'o}sp{\'a}l}, {Matter}, {van Boekel}, {Henning}, {Jaffe},
  {Juh{\'a}sz}, {Lopez}, {Menu}, {Mo{\'o}r}, {Mosoni}, \& {Sipos}}]{Varga2018}
{Varga}, J., {{\'A}brah{\'a}m}, P., {Chen}, L., {et~al.} 2018, \aap, 617, A83

\bibitem[{{Vigan} {et~al.}(2010){Vigan}, {Moutou}, {Langlois}, {Allard},
  {Boccaletti}, {Carbillet}, {Mouillet}, \& {Smith}}]{Vigan2010}
{Vigan}, A., {Moutou}, C., {Langlois}, M., {et~al.} 2010, \mnras, 407, 71

\bibitem[{{Vinkovi{\'c}} {et~al.}(2006){Vinkovi{\'c}}, {Ivezi{\'c}},
  {Jurki{\'c}}, \& {Elitzur}}]{Vinkovic2006}
{Vinkovi{\'c}}, D., {Ivezi{\'c}}, {\v Z}., {Jurki{\'c}}, T., \& {Elitzur}, M.
  2006, \apj, 636, 348

\bibitem[{{Zhu}(2019)}]{Zhu2019}
{Zhu}, Z. 2019, \mnras, 483, 4221

\end{thebibliography}

\begin{appendix}

\section{Observing setting and post-processing techniques} \label{Post_proc}
\begin{figure*}
  \centering
 \includegraphics[width=17cm]{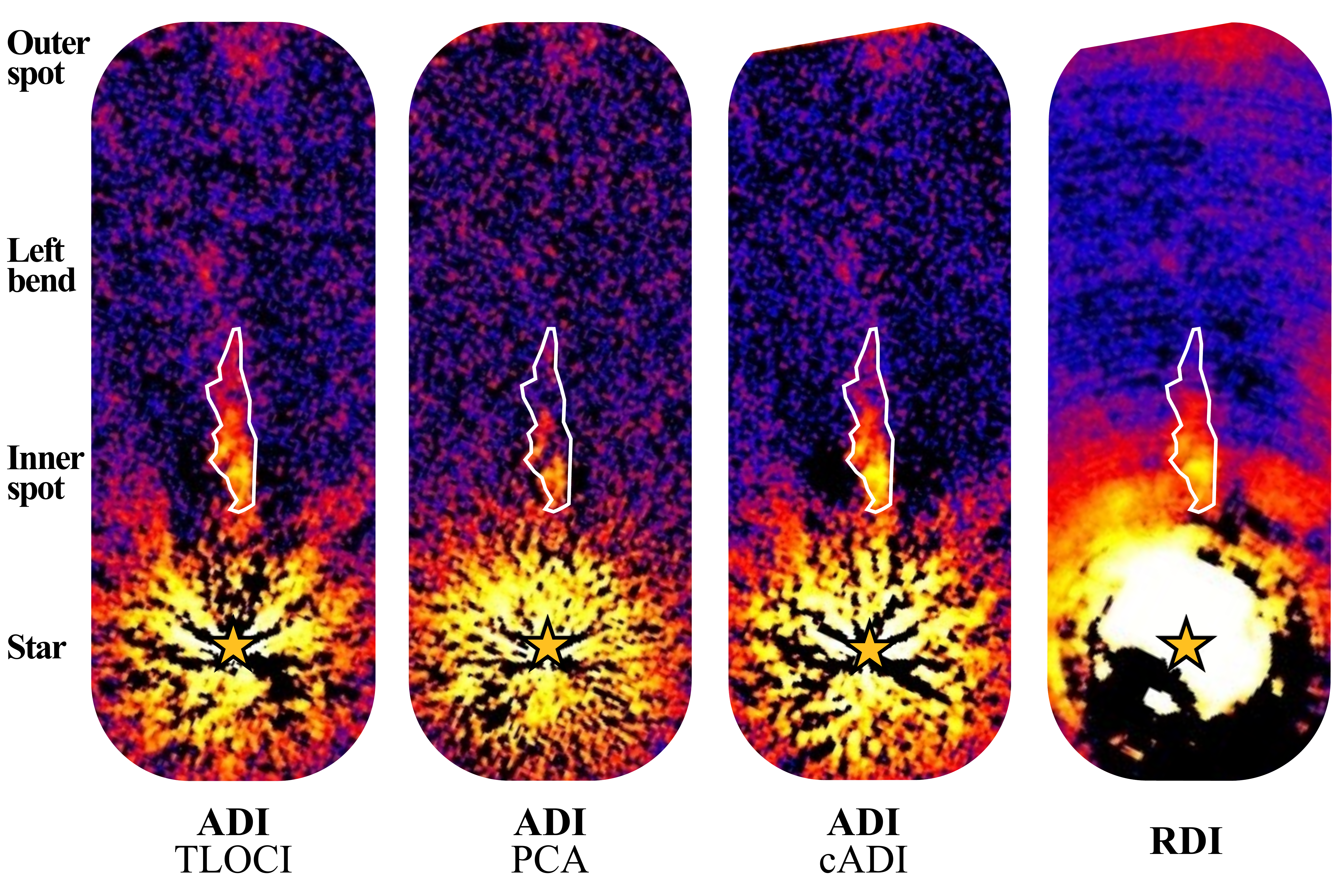} 
  \caption{Comparison of different post-processing techniques on the IRDIS H3 images from 2017. The first three figures are obtained in ADI, the fourth in RDI (see Sect.\,\ref{Observations}). {The strong quasi-centrosymmetric signal is the stellar continuum residual.} The jet morphology discussed in Sect.\,\ref{Jet_detection} looks reasonably similar {(see the white line encircling the inner spot)} and all the features that we discussed in the text are always present (with the possible exception of the left bend in the PCA image). Images are rotated by 65\degree counterclockwise. The color scale is logarithmic and relatively arbitrary.}
          \label{IRDIS_postproc}
  \end{figure*}

Figure \ref{IRDIS_postproc} is a comparison of different post-processing techniques on the IRDIS H3 images from 2017. It is clear that all images look reasonably similar, motivating our choice of Sect.\,\ref{Jet_detection} to focus on one image only. In particular, in the RDI image all features have lower contrast than in the ADI images. Their morphology is however more reliable since this technique does not suffer from azimuthal self-subtraction as the ADI. The similarity with the ADI images suggests that the ADI bias on the jet morphology is limited {even in the azimuthal direction}. 

\section{Stellar properties of RY Tau} \label{Spec_type}
\begin{figure*}
  \centering
 \includegraphics[width=9cm]{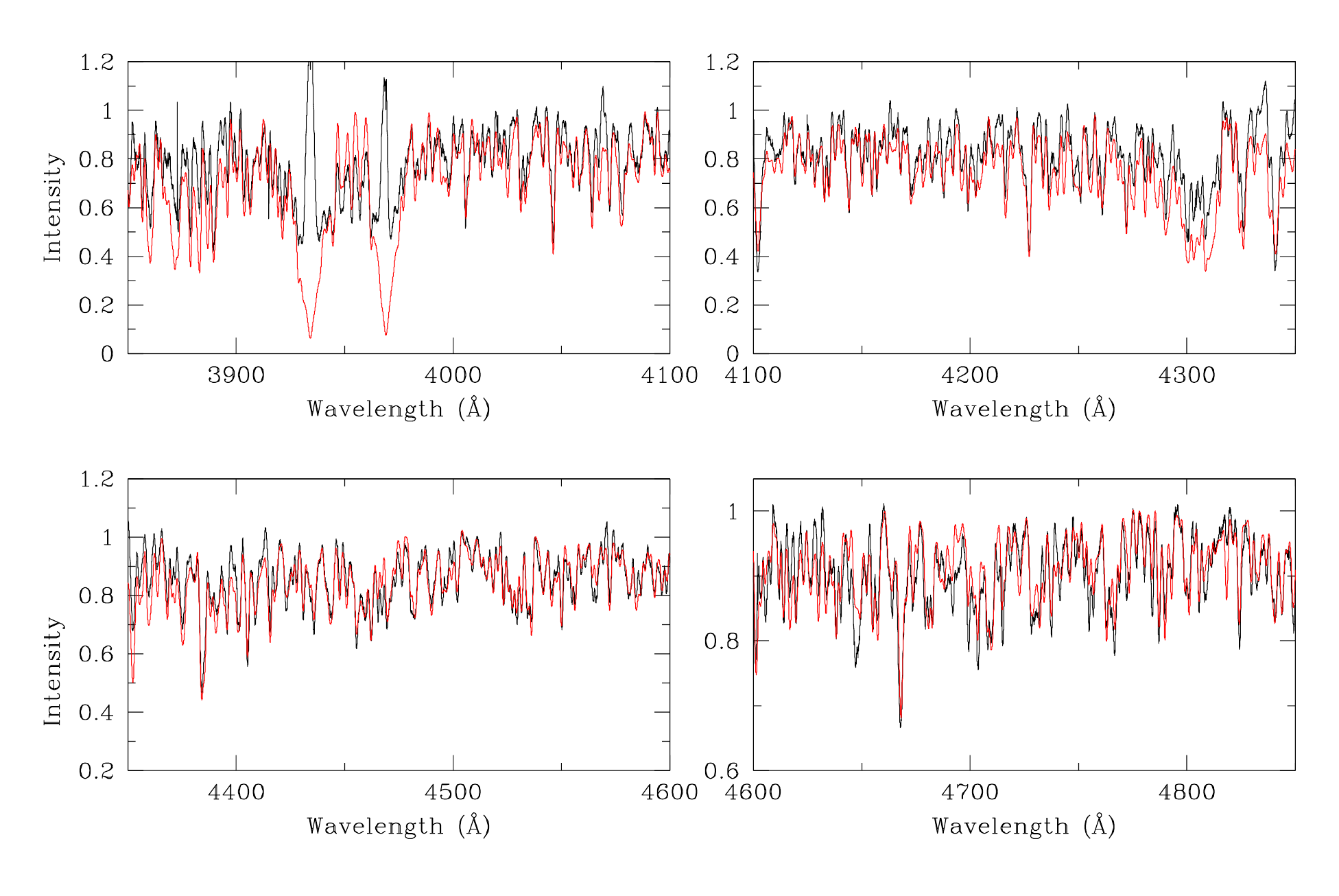} 
\includegraphics[width=9cm]{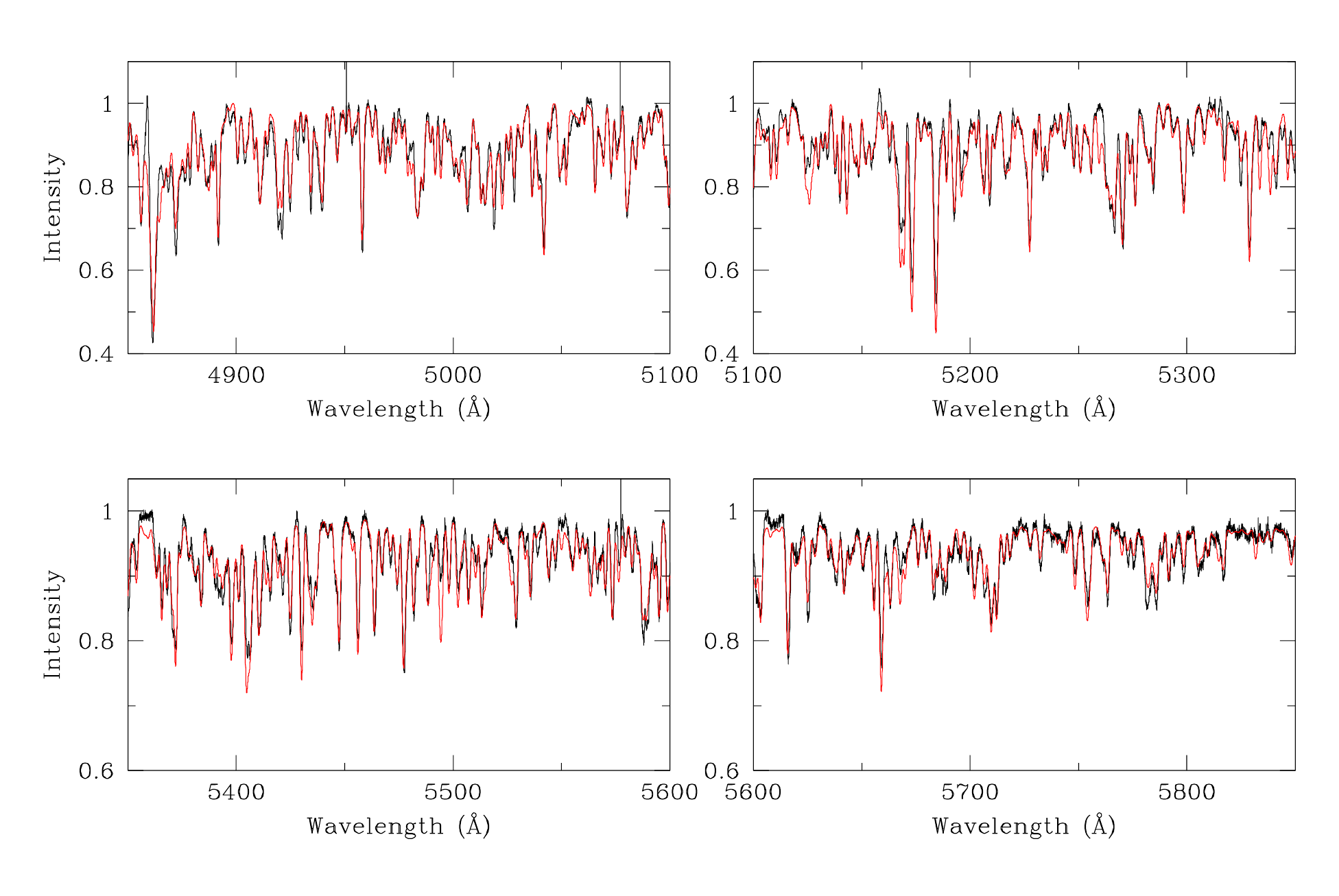}  
\includegraphics[width=7cm]{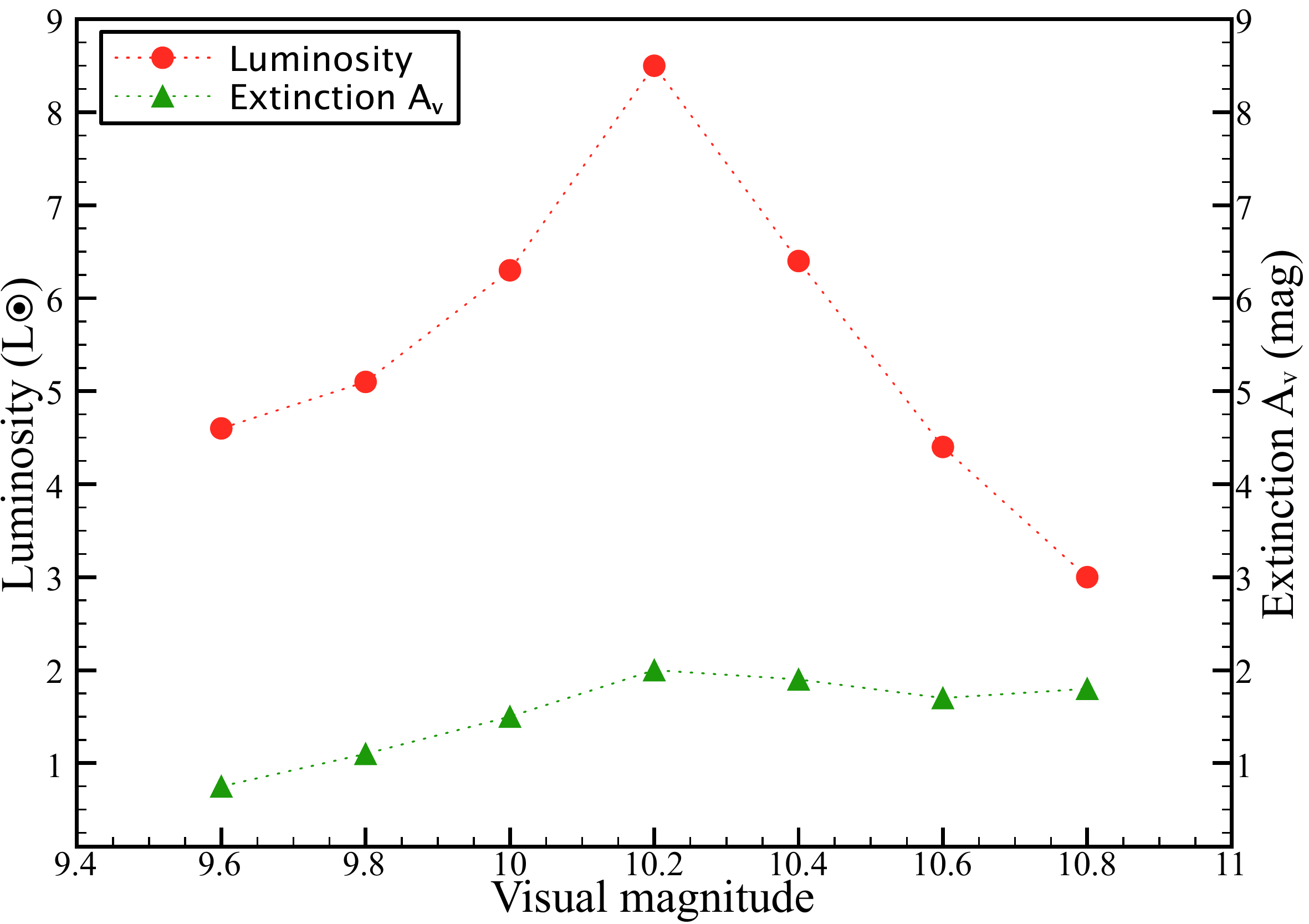}  
\includegraphics[width=7cm]{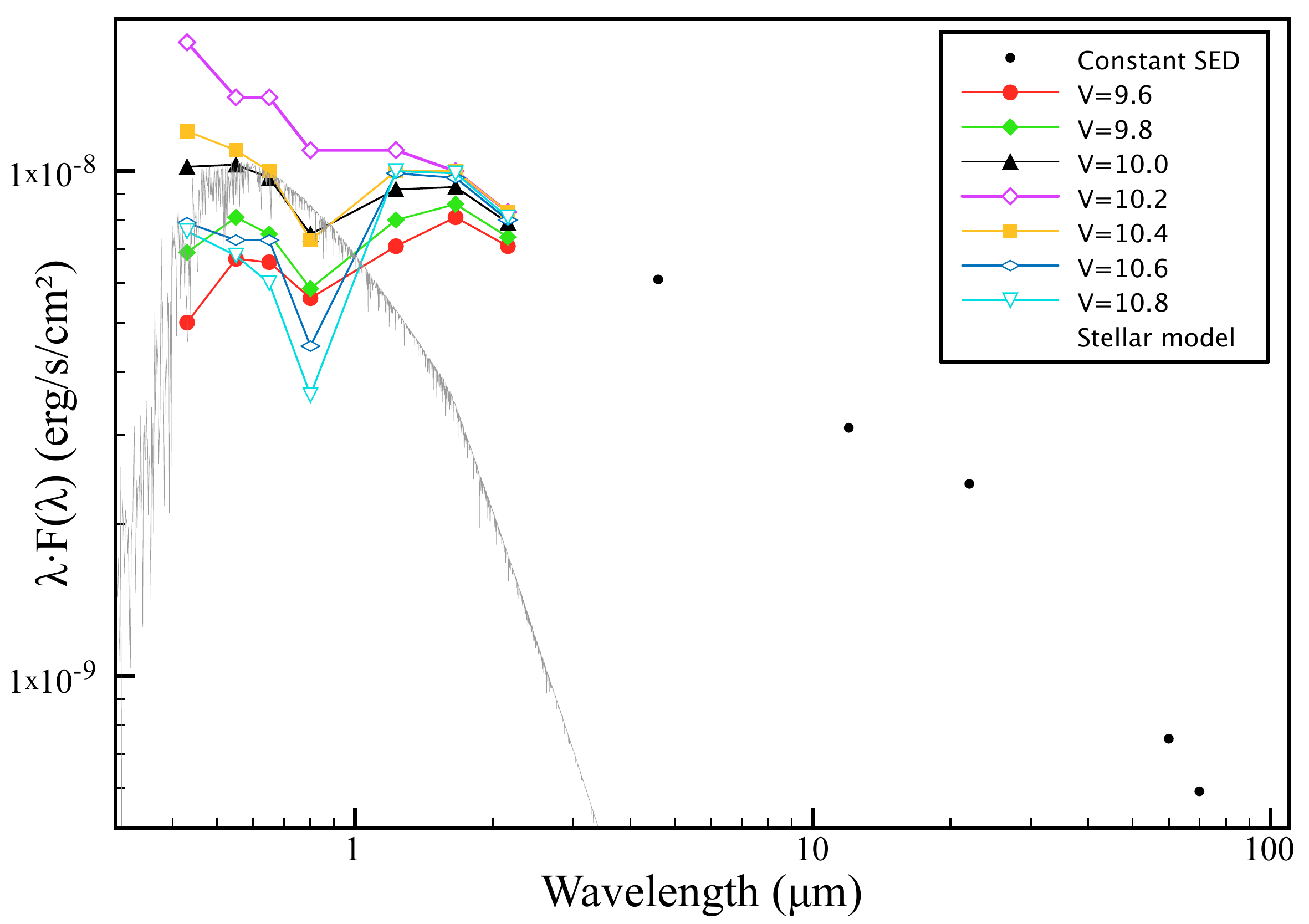}  
  \caption{Study of the stellar properties of RY Tau. The top panels are the optical EXPORT spectrum (black) compared to a synthetic model (red) with $T_{\rm eff}=5750$ K, log($g$)=3.58, and $v\cdot \sin(i)=52$ km/s. The third panel shows the different values of optical extinctions $A_{\rm V}$ and stellar luminosities obtained by adopting different brightnesses in the $V$ band as photospheric level. The fourth panel is the de-reddened SED of the source with different values of $V$ mag, and thus $A_{\rm V}$. A reasonable fit to the optical photometry and agreement with the NIR photometry is obtained only with $V=10.0$ and $V=10.4$. The stellar model is obtained for $V=10.0$ and $A_{\rm V}=1.5$ and corresponds to a stellar luminosity $L_*=6.3\,L_{\odot}$.}
          \label{Spectral_type}
  \end{figure*}

In Fig.\,\ref{Spectral_type}, we show the optical EXPORT spectrum used to constrain the effective temperature of RY Tau (see Sect.\,\ref{Stellar_properties_RYTau}), as well as the different values of visual extinction $A_{\rm V}$, of stellar luminosities, and the SED obtained with different scaling $V$ magnitude (see Sect.\,\ref{Stellar_properties_RYTau}).

\end{appendix}

\end{document}